# AI-Driven Test Case Generation from Natural Language Requirements: A Survey of Techniques and Research Gaps

Orimoloye Folorunsho, Hassan Reza
School of Electrical Engineering and Computer Science
University of North Dakota,
Grand Forks, ND 58201, USA
folorunsho.orimoloye@und.edu; hassan.reza@und.edu

***Abstract*** Software testing is critical for verifying that systems meet their specified requirements, yet it remains among the most time-consuming and expensive activities in development. While requirements-based test generation allows test cases to be derived early from requirements artifacts, generating them directly from natural language is challenging due to the inherent ambiguity and imprecision of NLP. Recent advances in AI, natural language processing (NLP), and large language models (LLMs) have made automating this pipeline increasingly feasible, but have also introduced new risks, including hallucination, reduced traceability, and inconsistent evaluation. This survey attempts to address four research questions: what AI and NLP techniques have been proposed for generating test cases from natural language requirements; what tools and frameworks support these approaches; how generated test cases are evaluated; and what research gaps remain. Following the systematic review guidelines of Kitchenham and Charters, we searched major scholarly databases spanning 2000–2025 and, after applying strict inclusion criteria, identified twenty-one primary studies through the systematic search; a targeted supplementary search added five further qualifying studies, yielding twenty-six primary studies for analysis. The literature is organized into three evolutionary eras, revealing that no existing approach simultaneously satisfies six key quality dimensions: automation, ambiguity handling, domain applicability, traceability, evaluation thoroughness, and hallucination control. The survey makes four main contributions: a three-era evolutionary synthesis of AI-based test generation; a related-work summary table comparing approaches, techniques, inputs, outputs, assumptions, and limitations; a six-criteria gap analysis showing that no current approach fully addresses all quality dimensions; and four actionable research guidelines targeting hallucination, traceability, complexity sensitivity, and compliance. Together, these position this work as a focused, LLM-era-inclusive synthesis of test-case generation from natural-language software requirements.

## I. INTRODUCTION

The demand for higher software quality and better customer experience, together with the growing complexity of modern systems, underscores the critical role of software testing in contemporary engineering practice. This need is especially evident in safety-critical domains such as aviation, automotive, finance, healthcare, and aerospace, where testing workloads are substantial, and defects can have severe consequences. Organizations must therefore ensure not only that software behaves as intended, but also that it does not behave in unintended ways or produce unintended outcomes. Software testing remains one of the most widely used means of assessing system reliability, quality, and functionality [36]. In a typical software project, approximately 50 percent of elapsed time and more than 50 percent of total development cost are devoted to testing activities [36]; in many projects, more than half of the overall schedule is now allocated to testing [37].

The software testing process is commonly organized into three stages: test case generation, execution, and evaluation [22]. Among these stages, test case generation is widely regarded as the most challenging and labour-intensive [22], [31]. During this phase, testers construct test cases to verify that functional requirements and design specifications are satisfied [31]. Although test cases may be created manually or generated automatically with tool support, manual construction is tedious, error-prone, and often inconsistent across projects and teams [4]. As a result, automating test case generation has become increasingly important in agile and continuous integration environments, where it helps reduce costs and effort while improving consistency [4].

Requirements-based test generation (RBTG) addresses this challenge by deriving functional test cases directly from requirements specifications or design artifacts rather than from source code [22]. Applying RBTG early in the software development life cycle enables testers to detect ambiguities and irregularities in requirements before defects propagate to later stages [1], [22]. However, RBTG also faces inherent limitations. Although natural language is widely used to document software requirements because of its accessibility and expressiveness [1], [39], it is often ambiguous and imprecise [24], [39]. This ambiguity complicates the direct automation of test-case generation from natural-language requirements [22].

In recent years, advances in artificial intelligence (AI) and natural language processing (NLP) have significantly expanded the potential to automate software testing. AI and NLP-based techniques, including syntactic and semantic analysis, information extraction, and learning-based reasoning, enable free-form requirement statements to be transformed into structured testing artifacts. For example, a requirement such as “the system shall display an error message if a user enters an invalid phone number” can be decomposed into test scenarios with explicit actors, preconditions, actions, and expected results [24]. Prior studies show that AI-driven test generation can accelerate testing, reduce manual effort, detect inconsistencies and anomalies, improve test procedures, and yield more effective test case designs [24], [25]. Likewise, NLP-based approaches can improve communication between developers and testers while supporting earlier detection of ambiguity and inconsistency in requirements [24], [27].

Among these advances, NLP plays a foundational role by providing the mechanisms for transforming free-form requirements into structured, testable representations. Natural-language requirements are among the most accessible yet most ambiguous artifacts in the software development life cycle. Consequently, each stage of the test case-generation pipeline, from identifying actors and actions to extracting conditions, resolving coreferences, and mapping requirement structure to test case fields, depends on recovering linguistic structure from free-form text. NLP provides the operational basis for this recovery. In particular, part-of-speech tagging and dependency parsing identify syntactic relations; semantic role labelling and named-entity recognition expose actor-action-object triples that map naturally to test case fields; and transformer-based contextual embeddings enable systems to generalize beyond fixed templates to the varied phrasings found in industrial requirements. The primary studies examined in this survey advance the field along one or more of these dimensions. For this reason, an NLP-focused perspective, rather than a purely model-driven or formal-specification-driven one, provides the most appropriate organizing framework for this survey.

However, the recent shift from traditional NLP pipelines to transformer-based LLMs has changed not only the scale of automation achievable but also the nature of the risks involved. Transformer-based large language models (LLMs) have substantially improved the semantic interpretation of natural-language requirements, but they have also introduced new risks, particularly hallucination, in which generated test cases are fluent yet semantically inconsistent with their source requirements. Li et al. [53] quantified this phenomenon in the HaluEval benchmark, reporting that 977 of 5,000 annotated ChatGPT responses (19.5 percent) contained hallucinations. Arora et al. [15] examined this risk directly through retrieval-augmented generation, providing the clearest primary study evidence to date of an architectural mitigation strategy. This emerging trustworthiness concern motivates Gap G1 (the absence of Hallucination Detection in LLM-Generated Test Cases), which is examined in this survey. More broadly, these developments expose limitations in the existing secondary literature, which has yet to provide a focused synthesis of AI/NLP-driven test case generation from natural-language requirements in the LLM era.

Against this background, this survey differs from prior reviews in four main respects. First, in terms of scope, Yang et al. [22] survey the broader requirements-based test generation literature across 267 primary studies, in which formal specification, model-based, and search-based approaches predominate. By contrast, this survey applies strict inclusion criteria (I4: natural-language input; I5: empirical AI/NLP test case-generation validation) to isolate the AI/NLP-driven subset that is not separately resolved in Yang's broader taxonomy. Second, in terms of temporal coverage, Boukhlif et al. [24] and Garousi et al. [27] examine only pre-LLM NLP-based testing, whereas Mustafa et al. [23] and Ahsan et al. [40] cover the 2000-2018 and 2005-2014 periods, respectively. This survey extends through October 2025 and, to the best of our knowledge, is the first to integrate primary studies from the LLM and hybrid eras alongside earlier rule-based and statistical-NLP work. Third, in terms of analytical focus, Wang et al. [25] survey LLM-based software testing broadly and treat natural-language input as only one of several input modalities. In contrast, this survey focuses specifically on the natural-language-to-test-case pipeline, where the risk of hallucination and the tension between traceability are especially pronounced. Fourth, in terms of contribution, prior surveys mainly provide taxonomies and classifications. In contrast, this survey presents a six-criterion gap analysis showing that no existing approach satisfies all six dimensions simultaneously and derives four actionable guidelines with explicitly defined deliverable artifacts. Taken together, these distinctions position this survey as the first AI/NLP-focused, LLM-era-inclusive, and gap-driven treatment of test case generation from natural-language requirements.

Motivated by these gaps in the existing review literature, this paper presents a focused survey of AI and NLP-based test case generation from natural-language software requirements, spanning early rule-based NLP systems, hybrid model-driven approaches, and modern transformer-based LLM techniques. The main contributions of this work are as follows: (C1) a three-era evolutionary synthesis grounded in twenty-six confirmed primary studies; (C2) a Related Work Summary table comparing approaches, techniques, inputs and outputs, assumptions, and limitations; (C3) a six-criteria comparison and gap-analysis table showing that no existing approach simultaneously satisfies all six quality dimensions; and (C4) four concrete, actionable research guidelines addressing the identified gaps.

To support this objective, the remainder of this paper is organized as follows. Section II reviews background on requirements-based testing, NLP in software engineering, and the motivation for integrating AI techniques. Section III presents the research questions and the systematic literature review methodology, including the search strategy, inclusion and exclusion criteria, study selection, and quality assessment. Section IV analyzes traditional NLP-based test case generation approaches, whereas Section V examines AI-driven and LLM-based techniques. Section VI compares tools and frameworks using the Related Work Summary and Six Criteria Comparison tables. Section VII identifies four research gaps (G1–G4) with empirical support from the primary studies. Section VIII proposes four actionable guidelines (R1–R4) that address these gaps. Section IX concludes the paper.

## II. BACKGROUND

Following the motivation, scope, and contribution established in the Introduction, this section presents the conceptual foundation for the subsequent analysis. It first examines requirements-based test case generation as the immediate problem context, then considers the role of natural language processing in software engineering as the enabling technical basis and finally highlights the emergence of artificial intelligence and large language models as a significant shift in the field. Together, these perspectives show the progression from formal, rule-based approaches to contemporary AI-driven methods for generating test cases from natural language requirements.

### A. Requirements-Based Test-Case Generation

Requirements are central to software development because they specify functionality, usability, performance, reliability, security, and design constraints [22]. They serve as a blueprint for both developers and testers, promoting clarity and reducing miscommunication across the development life cycle [22], [23]. Consequently, the quality of software development and testing depends heavily on the quality of the underlying requirements [23]. Requirements-based testing derives test cases from requirements while abstracting the implementation's internal structure [41]. In this context, requirements are categorized as functional, non-functional, or domain specific [40]. Functional requirements describe what the system must do, non-functional requirements define quality attributes and constraints, and domain requirements capture application-specific properties that may be functional or non-functional.

Bruel et al. [39] identified five main approaches for expressing software requirements, which are distinguished by the primary notation or style used in specifying the requirements:

- Natural language is a fully informal approach that expresses requirements using English or other human languages.
- Semi-formal approaches combine natural language with structured modelling elements, such as the Systems Modelling Language (SysML) [42] and the Unified Requirements Modelling Language (URML) [43].
- Seamless approaches integrate requirements directly with implementation and design through programming language constructs [39], [44], exemplified by Seamless Object-Oriented Requirements (SOOR) [44].
- Automata and graph-based approaches rely on graph-theoretical representations [6] to specify requirements, including UML activity and sequence diagrams [45], labelled transition systems [45], and finite-state machines [46].
- Mathematical approaches employ formal notations [22] or model-checking-based methods [47] to develop specifications and generate test cases, using formal specification languages such as Z [47], [50] and the Structured Object-Oriented Formal Language (SOFL) [49], [48], the latter exemplified by tool support for SOFL-driven automatic test case generation and result analysis.

These requirement expression styles differ substantially in their suitability for requirements-based test generation (RBTG). Mathematical, automata-based, and seamless approaches provide formally structured representations that support deterministic test case derivation, but they impose high modelling overhead and require specialized expertise. As a result, their adoption is largely concentrated in safety-critical or highly regulated domains. By contrast, natural language and semi-formal requirements dominate industrial practice because they are more accessible and flexible, despite their inherent ambiguity and imprecision. This tension has increasingly shifted RBTG research toward bridging the gap between informal or semi-structured natural-language requirements and executable test artifacts, thereby motivating the use of natural language processing, machine learning, and, more recently, large language models to recover structure, intent, and testable semantics from free-form requirement text.

### B. Natural-Language Processing In Software Engineering.

Manual testing is often costly and labor-intensive. To mitigate these challenges, NLP techniques such as language modelling, named-entity recognition, and text embedding enable advanced automated text processing, which can substantially reduce testing effort and cost [51]. More broadly, NLP is grounded in decades of research on text mining, statistical language modelling, and information extraction. Kao and Poteet [38] provide a foundational reference covering these core NLP and text-mining techniques, many of which underpin modern software engineering applications.

In recent years, software engineers have increasingly applied NLP to analyze human language and design artifacts and to automate tasks such as test case generation [5], [7], [51]. Zhao et al. [28] systematically mapped 404 studies on NLP for requirements engineering and showed that techniques for requirements analysis, extraction, and classification form a direct technical substrate for the downstream test-case-generation pipelines examined in this survey. NLP converts free-form text into structured representations through techniques such as tokenization, part-of-speech tagging, syntactic parsing, semantic role labeling, and coreference resolution [24]. For example, Viggiato et al. [51] used tokenization to decompose requirement sentences into word sequences, which were then reduced to their root forms through lemmatization. Similarly, Carvalho et al. [3] and Verma and Beg [2] showed how parse trees and controlled natural languages can be used to map requirement sentences directly to test case structures.

Early approaches in this area relied primarily on rule-based parsers. In contrast, more recent work has shifted toward machine-learning pipelines built on models such as conditional random fields, recurrent neural networks, and transformers [20], [24]. This evolution reflects a broader move toward data-driven and representation-learning approaches that can better accommodate the linguistic variability and contextual complexity of natural language.

### C. Motivation for AI Integration

Traditional NLP techniques have long been applied to software engineering tasks, including software testing. Although these methods are effective for capturing syntactic structure and limited semantic information, they are less effective for domain adaptation, contextual reasoning, and implicit requirement semantics. Building on this foundation, machine-learning techniques and, more recently, large language models (LLMs) have substantially expanded these capabilities by learning mappings between natural language requirements and executable test artifacts [20], [25].

ML and LLM-based approaches can automatically identify edge cases, constraints, and negative scenarios, all of which are essential in Agile and DevOps environments [35]. In addition, LLMs can interpret paraphrases, reason over contextual variation, and generate executable code for test scripts [20], [33]. These advances position AI-driven techniques as a natural progression beyond rule-based and feature-engineered NLP pipelines. Tertiary evidence further supports this shift. Amalfitano et al. [29], in a tertiary study of twenty secondary reviews, confirm that the use of artificial intelligence in software testing is now well established. However, they also observe that test case generation remains under-mapped at the secondary study level, thereby reinforcing the need for the focused synthesis undertaken in this survey. Operational precedents further illustrate both the promise and the limitations of current approaches. Sawada et al. [30] demonstrate intelligent requirements-to-test traceability at NASA-JPL using NLP and ML techniques, providing concrete evidence that AI-driven traceability is feasible even in safety-critical settings. At the same time, their work shows that such capabilities are typically realized through customized pipelines rather than end-to-end generative systems. This distinction directly anticipates the traceability challenge later formalized as Gap G2 in this survey.

## III. Systematic Review Methodology

To ensure a comprehensive and unbiased review of the literature, we conducted a systematic literature review (SLR) following the guidelines proposed by Kitchenham and Charters [52]. This section outlines the review protocol, including the research questions, search strategy, study selection criteria, quality assessment, data extraction, and synthesis methods.

We developed the SLR protocol for this survey paper by clearly defining the research objectives. We formulated research questions focused on key aspects of AI-driven test-case generation from natural language requirements. We then used a comprehensive search strategy across multiple academic databases and digital libraries. To ensure only relevant, high-quality research articles were included, we applied specific inclusion and exclusion criteria. The study-selection process involved screening titles, abstracts, and full texts, followed by a thorough review using a quality-assessment checklist to evaluate methodological rigor and relevance.

For data extraction, we systematically recorded key information from each selected study, including the proposed techniques, the tools and frameworks developed, the evaluation metrics, and the identified gaps, using a standardized form. Next, during the synthesis phase, we organized and summarized the findings. This approach enabled clear comparison and identification of trends across the literature. By following this method, the review ensures transparency, reproducibility, and comprehensive coverage of the topic, ultimately providing a solid foundation for mapping the current state of the art and guiding future research directions.

### A. Research Goals and Objectives

The primary goal of this research is to systematically map and synthesize the current state of the art in generating test cases from natural language software requirements using artificial intelligence (AI) and natural-language processing (NLP) techniques. The objectives are:

- To identify and categorize the various AI and NLP techniques proposed for test case generation from natural language software requirements.
- To review and analyze the tools and frameworks available for supporting AI-based test case generation.
- To examine how generated test cases are evaluated and the metrics commonly used for their assessment.
- To highlight existing research gaps, limitations, and challenges in the field, thereby guiding future research directions.

These goals and objectives provide clear direction for the research, ensuring that the systematic review focuses on the most relevant aspects of AI-driven test-case generation from natural language requirements. The review aims to answer the following research questions (RQs):

**RQ1**: What techniques have been proposed for generating test cases from natural language software requirements using AI and NLP?

**RQ2**: What tools and frameworks have been developed to support AI-based test case generation from NLSRs?

**RQ3**: How are the generated test cases evaluated, and what metrics are used?

**RQ4**: What are the main research gaps and challenges in this area?

These research questions are operationalized by explicitly mapping them to the survey's evidence base. **RQ1** (techniques) is addressed in Sections IV and V through the corpus of twenty-six primary studies organized across the three evolutionary eras. **RQ2** (tools and frameworks) is addressed in Section VI, where Table III summarizes related work and provides a cross-cutting comparison of tools. **RQ3** (evaluation) is addressed in Section VI through Table IV, particularly criterion K5 on evaluation thoroughness. **RQ4** (gaps and challenges) is addressed in Section VII through four formalized research gaps: G1 (hallucination), G2 (traceability), G3 (complexity sensitivity), and G4 (compliance). Each gap is grounded in specific cell values in Table IV and in external quantitative baselines from [16] and [53] and is mapped one-to-one in Fig. 7 to the four actionable recommendations (R1-R4) developed in Section VIII. This explicit chain from research question to section, table, gap, and recommendation ensures that each finding remains traceable to its evidentiary basis.

These questions guide the search, selection, and synthesis of primary studies.

### B. Search Strategy

We conducted a systematic search of the following digital libraries and databases, covering the period from January 2000 to October 2025: IEEE Xplore, Engineering Village, ScienceDirect (Elsevier), ACM Digital Library, SpringerLink, arXiv.org (for recent preprints), and published books and other journals/databases.
We constructed search strings by combining keywords related to the core concepts, specifically test-case generation,

natural language requirements, and AI/NLP techniques. To ensure broad coverage, we applied Boolean operators and wildcards. The core search query was:
("Test Case Generation" AND "Artificial Intelligence" AND "Natural Language Requirements") AND ("natural language processing" OR "natural language processing systems" OR "software testing" OR "software engineering" OR "requirements engineering" OR "specifications" OR "large language models" OR "natural languages" OR "specification languages" OR "automatic test pattern generation")
We further refined the searches using specific technique names such as rule-based, UML, OCL, transformer, GPT, and performed backward and forward snowballing on key papers, including full inspection of the 324-reference list of Yang et al. [22].

### C. *Inclusion and Exclusion Criteria*

We included studies that met the following criteria:

- **I1**: Peer-reviewed journal articles, conference papers, or authoritative arXiv preprints.
- **I2**: Published between 2000 and 2025.
- **I3**: Written in English.
- **I4**: Focus on generating test cases (or test artifacts) from natural language requirements.
- **I5**: Use some form of AI or NLP techniques, for example, rule-based systems, statistical ML, deep learning, LLM, or hybrid approaches.
- **I6**: Present empirical results or a clear methodological contribution.

Studies were excluded if:

- **E1**: They were purely theoretical with no implementation or evaluation.
- **E2**: They focused on test execution or test prioritization without generation.
- **E3**: They used only formal specifications (e.g., Z, VDM) without natural-language input.
- **E4**: They were short papers, posters, or tutorials lacking sufficient detail.
- **E5**: They were duplicate publications of the same study (only the most complete version was retained).
- **E6**: They were published in venues not indexed in IEEE Xplore, ACM Digital Library, Scopus, Web of Science, or DBLP, or that exhibit characteristics associated with predatory publishing.

### D. *Study Selection Process*

The selection process followed the PRISMA (Preferred Reporting Items for Systematic Reviews and Meta-Analyses) guidelines. It included four stages:

- Identification: The search returned 539 records before duplicates were removed.
- Screening: Titles and abstracts were evaluated against the inclusion criteria. 403 records were excluded, leaving 136 potentially relevant papers.
- Eligibility: The full texts of 136 papers were reviewed for eligibility. One hundred twenty papers were excluded due to factors such as a lack of NLP focus, insufficient detail, failure to address test generation jointly under I4 and I5, or venue quality concerns under E6.
- Inclusion: A total of twenty-one primary studies were ultimately included in the review after strict application of I4, I5, and E6. Five of these studies were identified via backward snowballing on the reference list of Yang et al. [22]. In addition, a targeted supplementary search of the same digital libraries identified five further primary studies [56]-[60] that satisfy the same inclusion criteria (I4, I5) and venue-quality criterion (E6) but were not retrieved by the original systematic query; each was verified through full-text review. This yields a final corpus of twenty-six primary studies analyzed in the remainder of this survey.

Figure 1 presents the four-stage PRISMA flow diagram summarizing the filtration process: 539 records were identified, 136 full-text papers were assessed, and 21 studies were finally included. The left column shows the records retained at each stage; the dashed right column indicates how many records were excluded at that stage and the rationale for the exclusion. The green inclusion box anchors the final twenty-one primary studies that feed the remainder of this survey.

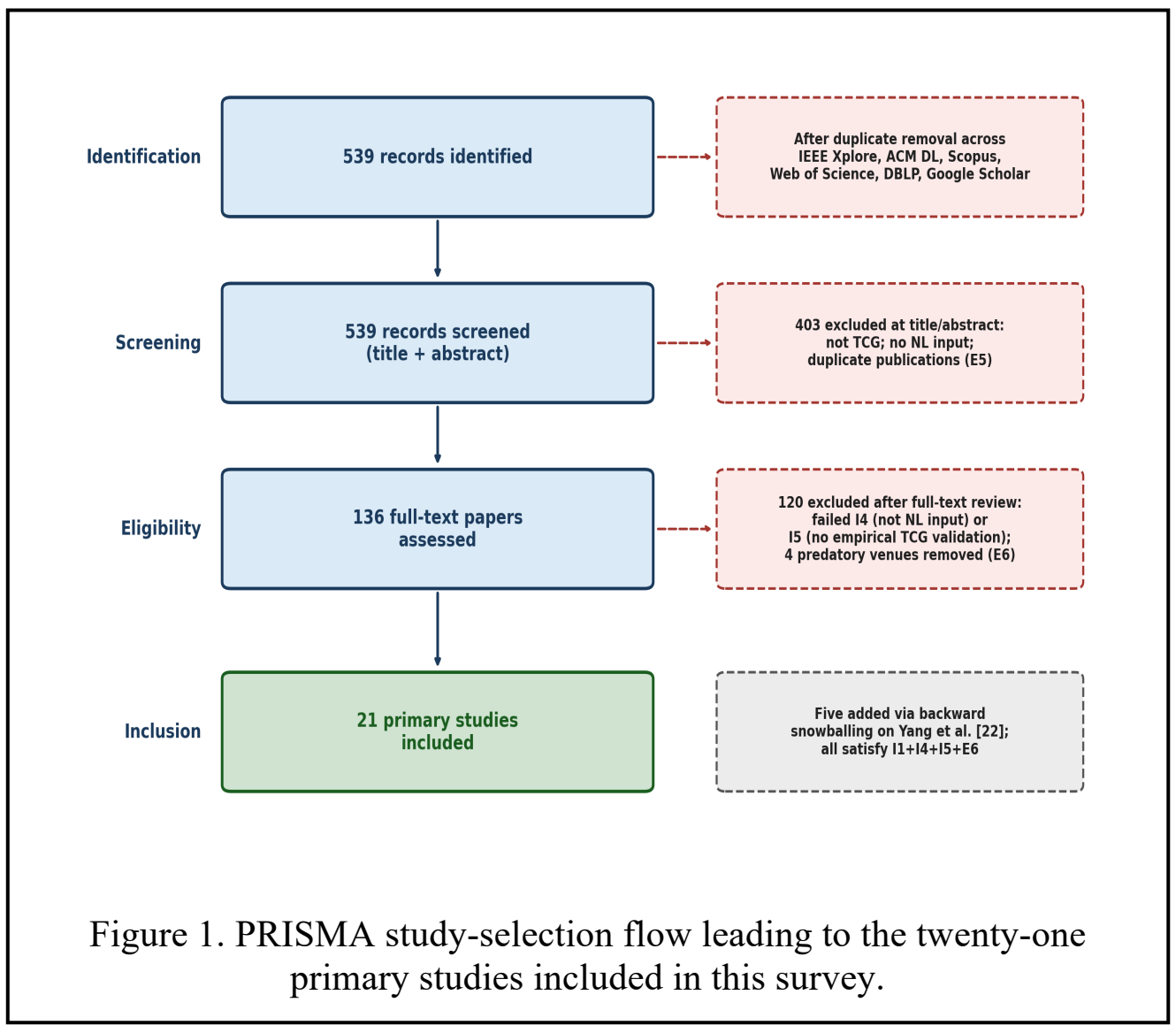

Figure 1. PRISMA study-selection flow leading to the twenty-one primary studies included in this survey.

### E. *Quality Assessment*

- Each included study was evaluated for quality using the following checklist adapted from [52]:
- QA1: Are the research objectives clearly stated?
- QA2: Is the proposed technique or tool described in sufficient detail?
- QA3: Is the evaluation thorough (e.g., use of benchmarks, comparison with baselines, statistical analysis)?
- QA4: Are the limitations and threats to validity discussed?

We concluded the assessment by rating studies on a three-point scale: yes = 1, partially = 0.5, and no = 0. The total quality score contextualized the findings without excluding any studies; all twenty-six confirmed primary studies met the minimum quality threshold of 2.0.

### F. *Data Extraction*

We created a data-extraction form to gather the following information from each primary study: bibliographic details (authors, year, title, venue); research objectives; input requirements type (structured, unstructured, domain-specific); NLP/AI techniques used; semantic representation (if any); output test artifacts (test steps, executable scripts, formal specifications); tooling context (prototype, open-source, commercial); evaluation metrics and results; application domain; and reported limitations and future work. We synthesized the extracted data through thematic analysis and categorized studies by technique era: rule-based, model-based, ML, and LLM/hybrid. We also grouped them according to taxonomy dimensions developed through the analysis. Using these categories, we identified patterns, trends, and research gaps based on limitations reported in the primary studies.

## IV. Traditional NLP-Based Test Generation

This section addresses RQ1 by examining traditional techniques for generating test cases from natural language requirements, including controlled natural languages (CNL), syntactic and semantic parsing, template and boilerplate approaches, and model-based methods grounded in UML and related formalisms. The discussion spans two historical phases: the Formalism Era (2000–2012) and the Statistical/NLP Era (2012–2022). Throughout this section, Figure 2 is used as a conceptual five-phase abstraction of requirements-to-test generation rather than a rigid workflow. Depending on the approach, phases may be emphasized unevenly, collapsed, specialized, revisited iteratively, or only partially realized.

The Formalism Era relied primarily on deterministic NLP pipelines, controlled natural languages, and direct translation from requirements into formal specifications. In contrast, the Statistical/NLP Era introduced greater flexibility through template matching, pattern extraction, machine-learning-based classifiers, and structured intermediate models, such as UML and Object Constraint Language (OCL) mappings.

Application of the strict inclusion criteria I4 (natural language input) and I5 (empirical validation for test case generation) yields no confirmed primary studies before 2012. This absence is methodologically meaningful rather than a limitation of the search. It reflects the dominant formal specification-first posture of the period, exemplified by approaches based on Z, SCR, VDM, B, and TLA+, which are excluded under I4 because they accept formal rather than natural language input. The encyclopedic model-based testing tradition of this era [32] and subsequent retrospectives on formalism in requirements engineering [39] confirm this orientation, while foundational software testing literature [31], [36] documents widespread reliance on deterministic processing, controlled natural languages, and specification-to-test-case translation pipelines.

Requirements-based test suite work during this period operated almost exclusively over formal representations, as illustrated by Vaysburg et al.'s dependence analysis approach to requirements-based test suite reduction [41] and Cartaxo et al.'s UML sequence diagram plus labelled transition system method [45]. Earlier work in the Z with Isabelle tradition [50], as well as later extensions applying Z-based model checking to interlocking systems [47], further exemplify the era's formalism-first stance.

Importantly, this tradition did not disappear after 2012 but was absorbed into hybrid pipelines that accept controlled natural language input while retaining formal specification translation as an intermediate step. NAT2TESTSCR [3], which uses SCR semantics, and NAT2TESTCPN [8], which targets Coloured Petri Nets, constitute primary study evidence of this bridge between controlled natural language and formal test case derivation. Broader coverage of pre-2012 work on test case generation from natural language lying outside the strict I4 + I5 scope of this survey is provided indirectly by related work mapping studies [23], [27], which reviewed this earlier literature under more inclusive criteria.

### A. *Controlled Natural Languages (CNL)*

Controlled Natural Languages (CNLs) represent one of the earliest and most influential approaches to reducing linguistic variability in requirements-based test case generation. By constraining natural language input with restricted grammars and vocabularies, CNL-based methods enable deterministic parsing and formal downstream processing. As a unifying conceptual model for requirements-to-test-case generation from natural language specifications, this survey adopts the refined five-phase framework shown in Figure 2: requirement ingestion, linguistic analysis, structured representation, test-case synthesis, and executable tests. CNL-based frameworks illuminate this abstraction especially clearly because they narrow the ingestion space to controlled natural language forms, reduce variability in linguistic analysis, and support explicit formal representations for traceable test derivation and, in some cases, execution-ready test artifacts; however, not every implementation materializes these phases as fully separate stages.

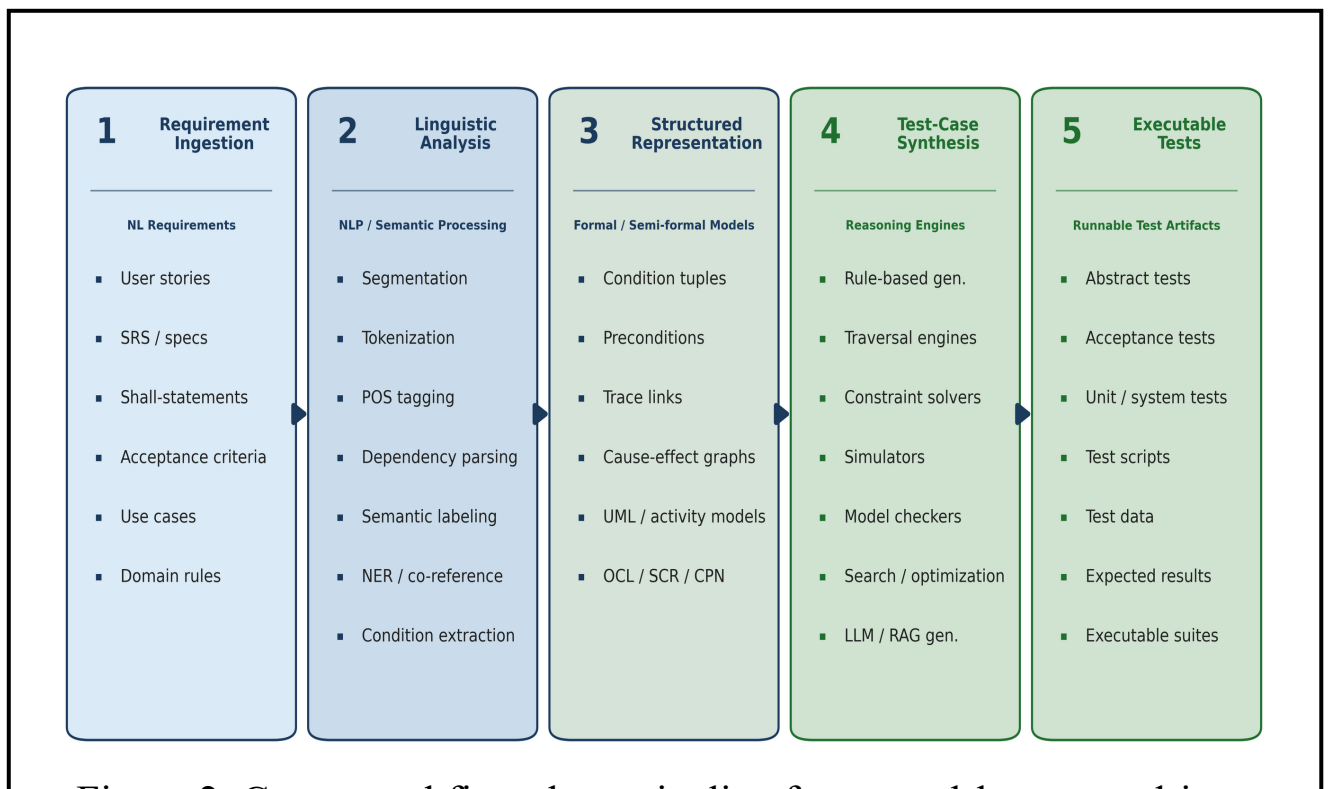

Figure 2. Conceptual five-phase pipeline for natural-language-driven test-case generation, from requirement ingestion to executable tests.

Viewed through the refined five-phase framework, NAT2TESTSCR derives its effectiveness from making the major requirements-to-test transformations explicit: controlled requirement ingestion via SysReq-CNL statements; lightweight linguistic analysis over restricted sentence patterns; structured representation in SCR tables; test-case synthesis via formal analysis with T-VEC; and the eventual production of execution-oriented test vectors. Figure 2 clarifies these recurring transformations and helps explain why CNL approaches achieve strong traceability and determinism at the cost of broader input flexibility. Depending on the approach, however, some phases may be collapsed, specialized, or revisited iteratively rather than implemented as strictly separate sequential steps.

Despite these strengths, NAT2TESTSCR requires authors to be trained in writing requirements within the controlled syntax, which limits scalability and adoption in environments where informal natural-language requirements dominate. This trade-off between precision and usability is a defining characteristic of CNL-based approaches and recurs throughout requirements-based test-generation research.

Silva et al. [8] extended this line of work with NAT2TESTCPN, published in the same journal in 2019, which translates the same CNL requirements into Coloured Petri Net (CPN) models and drives test case generation through simulation rather than constraint-based analysis. NAT2TESTCPN achieved approximately 80 percent mutation scores on the vending machine, nuclear power plant, and Embraer examples, and approximately 54 percent on the turn indicator example, with a peak mutation score of 85.07 percent on the Embraer case. In that example, NAT2TESTCPN matched NAT2TESTSCR's performance, while exceeding it on the remaining three systems, demonstrating that alternative formal back-ends can be effectively substituted within a CNL-driven pipeline.

Together, NAT2TESTSCR and NAT2TESTCPN provide strong primary-study evidence that controlled natural language approaches can enable highly effective, formally traceable test-case generation. At the same time, their reliance on constrained input languages underscores why subsequent research sought to relax linguistic restrictions through statistical NLP, machine learning, and hybrid approaches, as examined in the following subsections.

### *B. Syntactic and Semantic Parsing*

Whereas CNL-based methods reduce ambiguity by constraining how requirements are written, a complementary line of research seeks to recover structure directly from unconstrained natural-language requirements through syntactic and semantic parsing. Rather than imposing controlled grammars, these approaches apply linguistic analysis to identify entities, actions, conditions, and relationships in free-form text, thereby expanding applicability while accepting greater interpretive uncertainty.

Verma and Beg [2] developed an early method that combines part-of-speech (POS) tagging with parse-tree generation to transform requirement sentences into knowledge graphs representing entities and their relationships. Semantic frames were subsequently used to derive test inputs, conditions, and expected outputs from these graphs. Building on this foundation, later studies applied dependency parsing and semantic role labelling (SRL) to more reliably identify actor-action-object triples, thereby improving extraction accuracy and robustness [10], [12]. These parsing-based pipelines form the basis for many modern hybrid systems that combine linguistic analysis with downstream formal or generative test synthesis mechanisms.

Within the refined five-phase abstraction in Figure 2, these methods contribute most directly to the linguistic analysis phase through operations such as tokenization, parsing, semantic role labelling, named-entity recognition, and condition extraction, and to the structured representation phase through the construction of intermediate artifacts such as knowledge graphs, actor-action-object relations, and condition-bearing semantic structures. Their outputs typically support downstream test-case synthesis, but the extent to which synthesis and executable-test generation are integrated varies substantially across approaches.

While parsing-based techniques offer greater flexibility and broader applicability than controlled natural languages, their effectiveness depends on the accuracy of linguistic analysis and on the handling of the ambiguity inherent in unconstrained natural language. As a result, these approaches motivated subsequent research into template-based, statistical NLP, and learning-based methods that further relax input assumptions while maintaining testability and traceability.

### *C. Template* and Boilerplate Approaches

Between unconstrained parsing and fully controlled natural languages lies a more pragmatic design space: template and boilerplate approaches. Rather than enforcing a formal grammar, these methods guide requirements authors toward recurring syntactic patterns that reduce ambiguity while preserving some natural language flexibility. In this way, they trade part of the freedom of free-form text for greater regularity in downstream extraction and test case derivation.

Lim et al. [12] proposed a unified boilerplate approach that integrates Rupp's boilerplates with the Easy Approach to Requirements Syntax (EARS) to support test case information extraction from both positive and negative requirements. Their NLP-based system extracted actors, conditions, and responses from the PURE datasets [54], achieving overall correctness rates of 61.7 percent on Pointis, 50.0 percent on Mdot, and 10 percent on Npac. The authors report substantially lower correctness for negative requirements than for positive ones, highlighting a key limitation of template-guided extraction as linguistic complexity increases. While such approaches reduce ambiguity and improve consistency, they also constrain expressive flexibility and require users to learn and adhere to specific authoring conventions.

The industrial applicability of template-oriented workflows is demonstrated by SPECMATE [9], which automatically generates acceptance tests from structured acceptance criteria. When evaluated on 961 user stories in collaboration with Allianz Deutschland, SPECMATE automatically generated 56 percent of the 604 manually created test cases. Building on this line of work, Fischbach et al. [11] introduced CiRA (Conditionals in Requirements Artifacts). This RoBERTa-based NLP approach extracts conditional structures from informal natural language requirements and maps them to acceptance-test conditions via cause-and-effect graphs. In an industrial study involving Allianz Deutschland, Ericsson, and Leopold Kostal, CiRA automatically generated 71.8 percent of 578 manually created test cases and discovered 80 additional relevant cases that the manual testing process had missed.

Although template, boilerplate, and parsing-based approaches differ in how they recover structure from requirements, they converge on a common operational logic: linguistic elements in the requirement text are mapped to corresponding fields in the generated test case. Figure 3 visualizes this shared correspondence by showing how actors, actions, conditions, and expected system responses align with test-case fields such as actor role, test steps, preconditions, and expected results. For the example requirement "the system shall display an error message if a user enters an invalid phone number," the actor "user" maps to the Actor field, the action "enters an invalid phone number" maps to the Action/Test-Step field, the conditional clause maps to the Precondition/Trigger field, and the expected response "display an error message" maps to the Expected Result field. This mapping captures the common intermediate logic that systems such as SPECMATE and CiRA make explicit, and that rule-based extraction approaches such as Medeshetty et al. [17] exploit in a more implicit form.

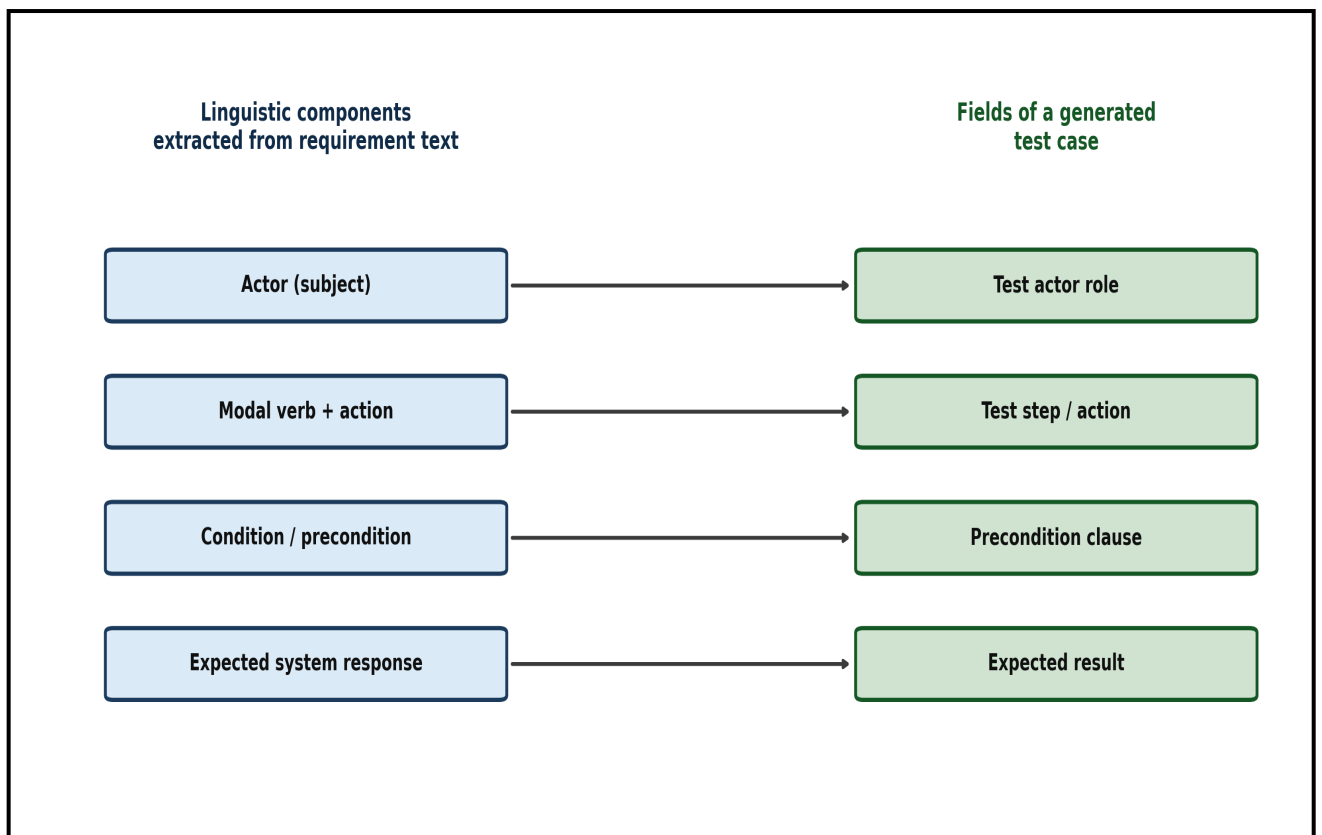

Figure 3. Conceptual correspondence between requirement-level linguistic elements and their mapped test-case fields across template-based, boilerplate-based, and parsing-based approaches.

Viewed through the refined five-phase abstraction in Figure 2, these approaches are best understood as concentrating on the boundary among requirement ingestion, linguistic analysis, and structured representation, where semi-patterned requirements are normalized, their conditions and responses are extracted, and the resulting intermediate forms support more reliable downstream test-case synthesis. In most cases, the final executable tests phase still depends on additional tooling or project-specific realization.

### D. Model-Based And UML Approaches

Where template, boilerplate, and parsing-based methods infer test case structure by mapping linguistic elements to corresponding test fields, model-based and UML-oriented approaches make that intermediate representation explicit by transforming natural language requirements into behavioral or structural models. Model-Based Testing (MBT), as defined by El-Far and Whittaker [32] in the Wiley Encyclopedia of Software Engineering, is the derivation of test cases from behavioral or structural models of the system under test. In requirements-based settings, these methods typically translate textual requirements into formal or semi-formal models, most Unified Modeling Language (UML) artifacts, before applying systematic test-generation techniques such as traversal, simulation, or constraint solving.

Wang et al. [10] proposed the Use Case Modelling for System Test Generation (UMTG) framework, which applies natural language processing to restricted use case specifications and translates them into UML models augmented with Object Constraint Language (OCL) constraints. Test cases are then generated by constraint-solving over the resulting formal representations. Evaluated in two industrial automotive case studies, UMTG achieved 95 percent correct translation of use case steps into OCL constraints and generated test cases that covered all expert-designed scenarios, as well as additional critical scenarios not previously considered. These results demonstrate the effectiveness of combining NLP-based extraction with formal modelling to achieve both automation and traceability.

Other model-based approaches similarly integrate linguistic preprocessing with behavioral modelling. Kamonsantiroj et al. [6] introduced a memorization-based method for generating test cases from concurrent UML activity diagrams, while Allala et al. [7] combined model-driven engineering (MDE) with NLP to transform natural language user requirements into test cases via model-to-model transformations. Keyword-based NLP techniques have also been applied to functional software requirements specifications to support UML-oriented test case generation [5]. In an industrial aerospace context, Olajubu et al. [4] translated natural language "shall" statements into a domain-specific modeling language, from which executable test cases were generated.

Within the refined five-phase abstraction in Figure 2, model-based and UML-driven approaches operate primarily across the structured representation, test-case synthesis, and executable tests portions of the framework. They use models such as UML, OCL, SCR, domain-specific languages, or other behavioral formalisms as the explicit bridge between natural-language requirements and systematic test derivation, while solvers, simulators, traversal engines, or model checkers support the generation and realization of concrete test artifacts.

Model-based approaches provide strong guarantees of coverage, determinism, and traceability, making them particularly attractive in safety-critical and regulated domains. However, they impose significant modelling overhead and often depend on restricted requirement formats or domain-specific assumptions. These constraints limit scalability and help explain why subsequent research has increasingly explored statistical NLP, machine learning, and large-language-model-based techniques that seek to reduce manual modelling effort while preserving testability.

Additional Statistical/NLP-era studies further confirm these patterns across new domains and intermediate representations. At the controlled-input end, Yue et al. [57] proposed Restricted Test Case Modeling (RTCM), a restricted natural-language notation coupled with the aToucan4Test tool and applied it to three subsystems of Cisco Systems Norway's video-conferencing system. The approach transformed nine test-case specifications into 246 automatically executable test cases, demonstrating industrial-scale traceability from constrained natural language to executable artifacts. In a model-based variant, Santiago Júnior and Vijaykumar [56] developed SOLIMVA, which translates natural-language requirements into Statecharts and derives test cases through the GTSC tool using combinatorial designs. The method was validated on the SWPDC software embedded in a satellite payload at Brazil's National Institute for Space Research (INPE), thereby extending the evidence base to the space domain. Other studies adopt alternative structured intermediates. Nayak et al. [59], for example, used a knowledge graph populated through named-entity recognition to generate test cases from domain-language requirements in an automotive project at Bosch, with the graph compensating for information absent from individual requirements. Similarly, Aoyama et al. [58] translated natural-language specifications into propositional-network and decision-table models to generate executable tests, together with a test-execution environment, for a heating, ventilation, and air-conditioning system developed under a software product line. Finally, Flemström et al. [60] mapped natural-language system-level requirements to passive test cases expressed as guarded assertions using the SAGA tool chain, including explicit extraction of temporal constraints, in a proof-of-concept study conducted at Bombardier Transportation Sweden. Collectively, these studies indicate that a central Statistical/NLP-era strategy—recovering structure from requirements through a controlled language, behavioral model, knowledge graph, decision table, or assertion form before test derivation—generalizes across aerospace, automotive, telecommunications, and embedded-systems domains.

Taken together, the approaches examined in Section IV demonstrate how deterministic NLP pipelines, controlled languages, templates, and model-based representations can recover structure from natural language requirements and support test case generation. However, these techniques typically rely on constrained input formats, predefined linguistic patterns, or explicit modelling effort, which limits their robustness and scalability when dealing with highly variable, informal, or evolving requirements. These limitations motivated a shift toward learning-based approaches that reduce manual authoring effort and operate directly on unrestricted natural language. The next section examines how machine learning and large language model-based techniques address these challenges by leveraging data-driven semantic understanding, while introducing new trade-offs in terms of traceability, determinism, and trustworthiness.

## V. AI-Driven and LLM-Based Techniques

Recent advances in artificial intelligence and natural-language processing have fundamentally redefined requirements-based automated test generation, shifting the field from rule and feature-engineered pipelines toward data-driven and generative architectures. This section addresses RQ1 and RQ2 by examining AI and NLP techniques proposed for generating test cases from natural language requirements, as well as the tools and frameworks that operationalize these techniques in practice. As in Section IV, Figure 2 is used here as a conceptual five-phase abstraction rather than a rigid workflow; depending on the architecture, phases may be fused within a single model invocation, externalized into retrieval or verification components, revisited iteratively, or only partially implemented. Against that framing, the discussion traces the evolution from pre-LLM machine-learning approaches to transformer-based and large-language-model (LLM) systems, highlighting improvements in coverage, precision, and adaptability when operating on complex, ambiguous requirement texts.

By linking techniques to concrete implementations, this section distils recurring strengths, limitations, and emerging architectural patterns. It begins with pre-LLM machine-learning systems, then turns to transformer and LLM-based approaches, before considering domain-specific applications and hybrid symbolic-neural designs that seek to recover some of the control and auditability lost in fully generative settings.

Building on this synthesis, Section VI systematically compares the identified tools and frameworks using uniform evaluation criteria, exposing unresolved tradeoffs and motivating the research gaps discussed later in the paper.

### A. *Pre-LLM Machine Learning Pipelines*

Between 2015 and 2020, AI-based test case generation from natural language requirements was dominated by traditional machine learning approaches that relied on supervised learning to map requirement sentences to structured test case components. Representative methods employed classifiers such as Support Vector Machines (SVMs) and Bidirectional Long Short-Term Memory networks (Bi-LSTMs) to perform sentence or fragment-level classification of requirement text into test artifacts [24], [26]. Compared with rule-based NLP systems, these approaches achieved greater flexibility and accuracy but relied heavily on labelled training data and exhibited limited cross-project generalization.

In Figure 2, these systems typically distribute effort across linguistic analysis, structured representation, and test-case synthesis, often via separate learned or feature-engineered components rather than a single end-to-end generator. Typical implementations combined linguistic feature extraction from requirements using part-of-speech tags and dependency relations, supervised classification of requirement fragments into test case elements such as inputs, actions, and expected outcomes, and sequence-oriented generation of structured test templates. While effective within narrowly defined training contexts, reliance on handcrafted features, annotated datasets, and project-specific models constrained scalability and hindered adoption across heterogeneous domains and evolving requirements.

These limitations exposed a structural ceiling for feature-engineered machine-learning systems. They created the conditions for transformer-based architectures, which absorbed more of the interpretation and generation burden into shared representations. The next subsection examines how that shift changed both the achievable level of automation and the nature of the associated risks.

### B. *Transformer and LLM Paradigm*

The 2020s marked the rise of transformer-based large language models (LLMs) capable of few-shot reasoning, significantly expanding the scope of requirements-based automated test generation. Transformer architectures have enabled models to capture long-range dependencies and contextual semantics in free-form requirement texts, allowing test cases to be generated directly from natural language inputs without the feature engineering and supervised classifiers required by earlier approaches.

Within the conceptual five-phase framework in Figure 2, transformer-based and LLM approaches often compress much of the linguistic analysis, portions of structured representation, and parts of test-case synthesis into a single generative inference process. At the same time, retrieval, prompting, verification, and test realization are frequently externalized into surrounding components, so these systems are better understood as reconfiguring the framework than as eliminating it.

Korraprolu et al. [18] evaluated six publicly available LLMs - BARD, ChatGPT-3.5, Claude, Gemini, GPT-4 Omni, and LLaMA 3 for generating test cases from English natural language requirements. Although all six models could produce executable tests, their performance varied considerably across coverage criteria, revealing differences in how effectively each model captured decision logic and conditional structure rather than surface-level syntactic completeness.

Figure 4 summarizes Korraprolu et al.'s benchmark across twenty-five requirements using four metrics: decision coverage, condition coverage, Modified Condition/Decision Coverage (MCDC), and execution rate. Gemini and ChatGPT-3.5 achieved the highest average decision and condition coverage, with Gemini reaching 85.2 percent and 82.9 percent, respectively, and ChatGPT-3.5 reaching 83.2 percent and 80.5 percent. Performance declined across all models on MCDC, where Gemini and Claude achieved the highest averages at 50.2 percent and 50.1 percent, respectively. By contrast, execution remained consistently high across all six models, with a median of 100 percent and an average above 96 percent. The central result is the large gap between execution and MCDC performance: the models generally produced runnable tests, but they were substantially less effective at covering interacting logical conditions. This divergence provides quantitative evidence that current general-purpose LLMs remain sensitive to the complexity of requirements, directly motivating Gap G3.

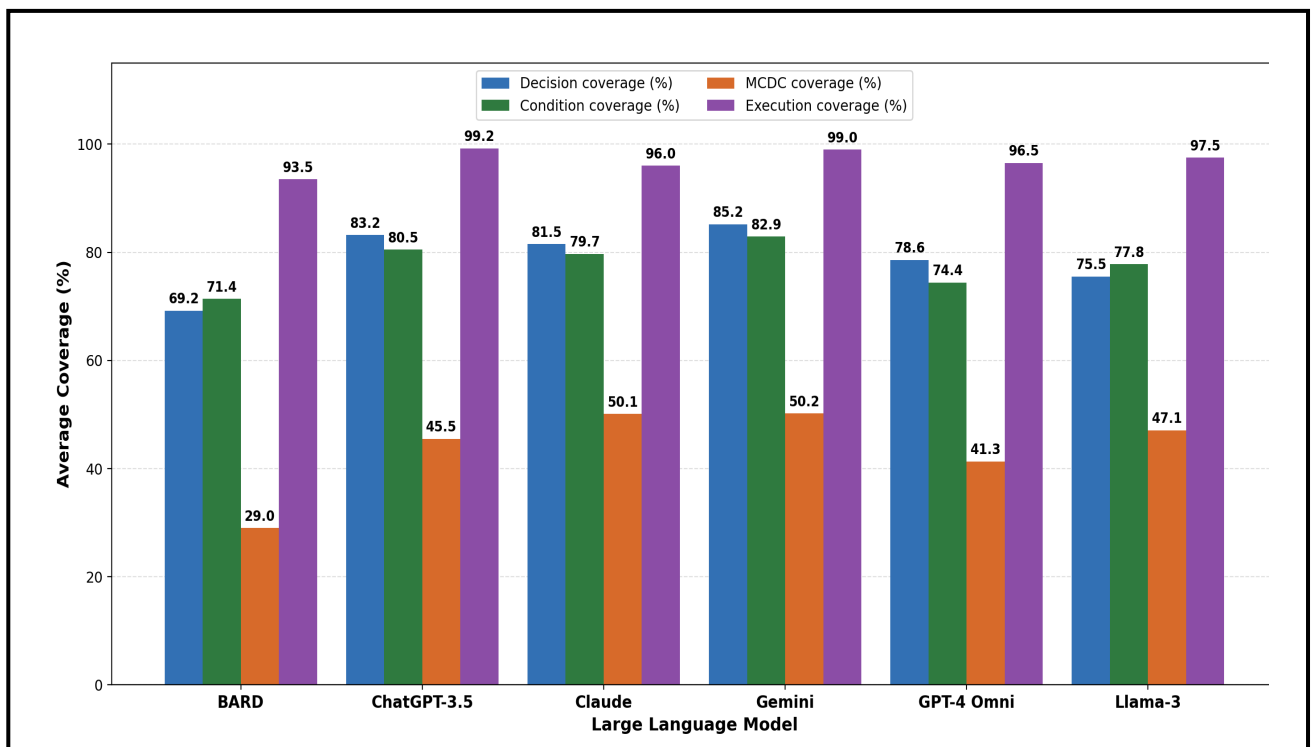

Figure 4. Average coverage of six general-purpose LLMs across decision, condition, MCDC, and execution criteria on twenty-five natural language requirements, based on the average values reported by Korraprolu et al. [18]. The six-model averages are 78.9 percent for decision coverage, 77.8 percent for condition coverage, 43.9 percent for MCDC, and 97.0 percent for execution. The marked execution-versus-MCDC gap indicates that these models usually generate runnable tests but remain weaker at exercising compound logical conditions.

This systematic degradation in MCDC performance as conditional interactions increase provides early empirical evidence that LLM-based test generation remains sensitive to requirement complexity, a limitation examined further in Gap G3.

To address these limitations in general-purpose models, Xue et al. [13] introduced LLM4Fin, a domain-specific LLM fine-tuned on corpora from financial regulation. By incorporating both explicit and implicit domain knowledge, LLM4Fin achieved a peak business-scenario coverage (BSC) of 98.18 percent on Dataset 3, with an average BSC of 91.89 percent across five datasets. In addition to improved coverage, LLM4Fin dramatically reduced test-generation time from approximately 20 minutes to an average of 6.99 seconds per generation. In contrast, the general-purpose ChatGPT model achieved only 49.65 percent average coverage and required approximately 19 minutes per generation.

Figure 5 compares LLM4Fin with four baselines - human experts, human nonexperts, general-purpose ChatGPT, and ChatGLM - using average business-scenario coverage (BSC) and generation time. The figure highlights two results. First, LLM4Fin substantially improves coverage over general-purpose ChatGPT, achieving an average BSC of 91.89 percent versus 49.65 percent. Second, it reduces average generation time from approximately 19 minutes to 6.99 seconds per case. Taken together, these results show that domain specialization can improve both effectiveness and efficiency in LLM-based test case generation.

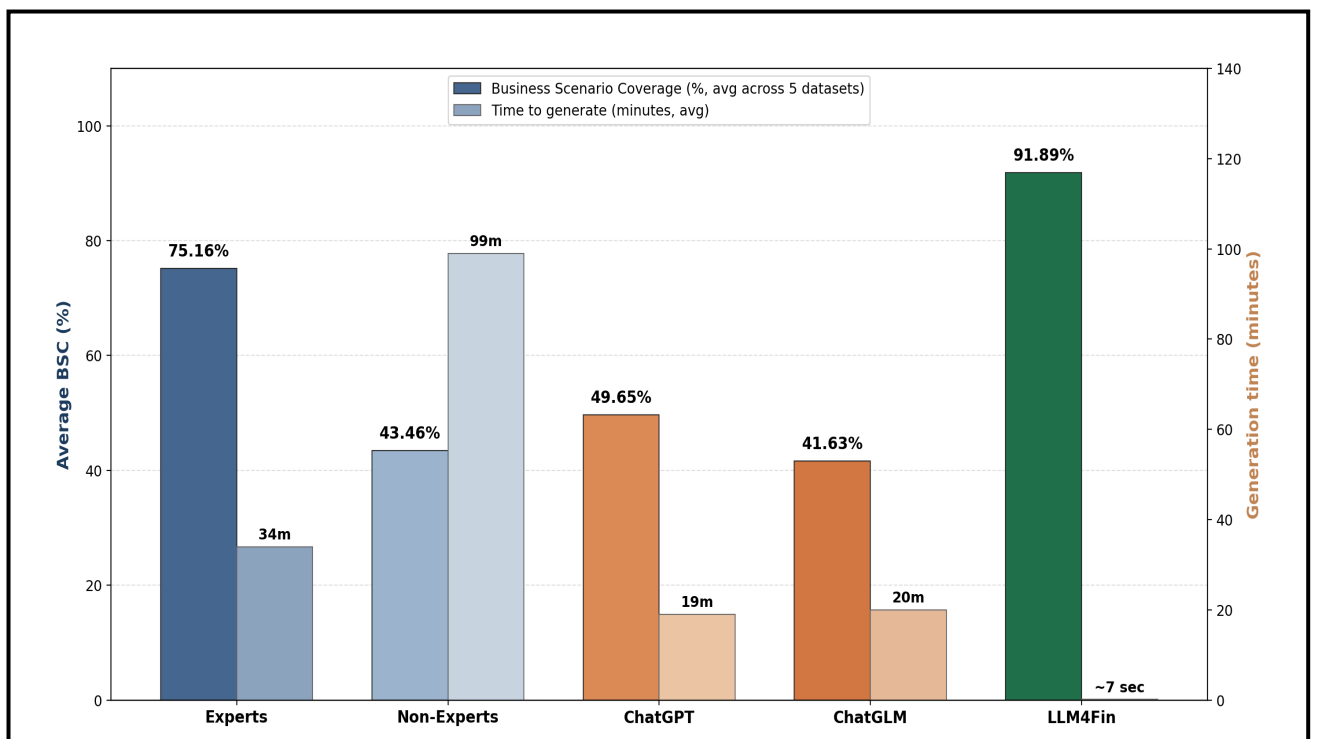

Figure 5. Average business-scenario coverage and generation time for LLM4Fin and four baselines, based on the average values reported by Xue et al. [13]. LLM4Fin achieves 91.89 percent average coverage and 6.99 seconds per case, compared with 49.65 percent and approximately 19 minutes for general-purpose ChatGPT.

Building on this line of work, Yin et al. [14] demonstrated the on-premises deployment of a 4-bit quantized LLaMA-3 70B Instruct model for quality-assurance teams, emphasizing secure test case generation without data leakage. Evaluating multiple prompting strategies, including zero-shot, one-shot, two-shot, and prompt-chaining, the authors reported 79.6 percent coverage for the base model (Table II) and up to 90.3 percent coverage for the optimal prompt-chaining plus two-shot sequential configuration (Table III). These results show that high-coverage LLM-based test generation can be achieved under enterprise data-privacy constraints.

Further optimization-oriented studies extend the applicability of LLMs to test generation. Alagarsamy et al. [20] introduced TestGPT, a GPT 3.5 model fine-tuned on a curated text-to-testcase dataset, achieving 78.5 percent syntactic correctness, 67.09 percent requirement alignment, 61.7 percent code coverage, and an 18.9 percent mutation score across five large open-source projects, substantially outperforming thirty-one other variants across eight baseline LLMs (basic GPT 3.5-turbo, Incoder, Starcoder, CodeT5, Bloom, CodeGemma, CodeLlama, and Gemini). Similarly, Takerngsaksiri et al. [21] proposed PyTester, which applies deep reinforcement learning to text-to-testcase generation and outperforms GPT-3.5 on the APPS benchmark.

Arora et al. [15] addressed trustworthiness through RAGTAG, an industrial retrieval-augmented generation framework evaluated at Austrian Post on two projects with bilingual German and English requirements. Generated test scenarios were assessed by four domain experts along five dimensions: relevance, coverage, correctness, coherence, and feasibility. Under the best-performing configuration, GPT-4.0 with few-shot prompting and single-chunk retrieval, the automatic scores were BLEU = 0.092, ROUGE = 0.419, and METEOR = 0.516. The low BLEU score is not surprising in this setting because test scenarios can admit multiple acceptable phrasings; accordingly, expert judgment provides the more informative evidence. The expert assessment found that RAGTAG generated scenarios aligned with the source requirements and feasible for practical testing. At the same time, the study indicates that current generative pipelines still depend on retrieval anchoring and human review, underscoring the absence of standardized hallucination-detection methods and formal traceability artifacts, thereby reinforcing Gaps G1 and G2.

Finally, Najmi and El-Dosuky [19] integrated ChatGPT with traditional NLP and deep learning components in a multi-stage framework for test case analysis, applying the approach to an optical character recognition system. Their study usefully bridges the purely pre-LLM and purely generative ends of the spectrum, thereby motivating the closer look at domain-shaped and hybrid architectures in the subsections that follow.

C. *Domain-Specific Applications*

If general-purpose LLM studies reveal the promise and instability of broad generative approaches, domain-specific applications show that performance can improve when requirements, domain knowledge, and evaluation settings are more tightly controlled. The studies in this subsection illustrate how architectural specialization can recover accuracy, speed, or robustness without reverting fully to the rigid constraints of earlier formal methods.

Medeshetty et al. [17] investigated domain-specific NLP for transforming natural-language automotive Electronic Control Unit (ECU) requirements into structured test-case specifications. Their study, conducted with Scania CV AB, combined rule-based information extraction with named-entity recognition (NER) and used 400 feature-element documents from Polarion, including a manually annotated subset of 200 entities for NER training and 10-fold cross-validation.

Table I summarizes the comparative performance of the rule-based and NER-based extraction methods across increasing levels of requirement complexity investigated by Medeshetty et al. [17].

TABLE I. REPORTED ACCURACY OF DOMAIN-SPECIFIC NLP EXTRACTION TECHNIQUES

| Method | Variant | Key Metric | Notes |
|---|---|---|---|
| Rule-based IE | Category 1 (single signal) | Accuracy = 95% | Baseline |
| Rule-based IE | Category 2 (≤ 4 signals) | Accuracy = 75% | Significant gain over manual [a] |
| Rule-based IE | Category 3 (> 4 signals) | Accuracy = 60% | Degrades with complexity |
| NER (SVM) | All categories | NER Acc = 77.3%; Prec/Rec/F1 = 0.816/0.773/0.770 | No significant gain over manual [a] |
| NER (Random Forest) | All categories | NER Acc = 40.8%; Prec/Rec/F1 = 0.263/0.408/0.316 | Substantially weaker than SVM |

a) **Statistical significance:** significance was tested against manual test-case writing using the Mann–Whitney U test, with a threshold of $p < 0.05$. The rule-based method significantly outperformed manual generation ($p = 0.037$), whereas the NER (SVM) method did not ($p = 0.806$).

From Table I, it can be seen that the rule-based extractor was strongly affected by requirement complexity, with accuracy declining from 95 percent for single-signal requirements to 75 percent for Category 2 and 60 percent for Category 3. Even so, it provided a substantial practical advantage over manual test-case generation, reducing effort from approximately 24 hours to 5 minutes per case; this improvement was statistically significant (Mann-Whitney U, $p = 0.037$).

Among the NER variants, the SVM model performed best, reaching 77.3 percent accuracy with precision, recall, and F1 scores of 0.816, 0.773, and 0.770, respectively, whereas the Random Forest variant achieved only 40.8 percent accuracy. Despite this competitive extraction accuracy, the NER approach did not yield a statistically significant improvement over manual generation ($p = 0.806$).

Taken together, these results indicate that deterministic rule-based extraction remains the stronger choice for well-structured automotive ECU requirements. In contrast, NER performance depends heavily on model selection and does not yet translate into a measurable end-to-end advantage over manual practice. This contrast motivates continued interest in hybrid rule-based and learning-based designs for more complex requirements.

A related pattern appears in another safety-critical domain. Shakthi et al. [16] combined NLP techniques, including NER, part-of-speech tagging, and dependency parsing, with few-shot and chain-of-thought prompting to generate test cases from ten satellite Functional Requirement Documents (FRD) at the U. R. Rao Satellite Center of the Indian Space Research Organization. Four LLMs were evaluated on the same task: GPT-4 generated 9 of 10 correct test cases (90 percent), Gemini-Pro generated 6 of 10 (60 percent), Llama-2-70B generated 4 of 10 (40 percent), and Mistral-7B generated 2 of 10 (20 percent). This stratified result provides clear quantitative evidence of model-dependent variation in NL-to-test generation accuracy under a common evaluation setting.

D. *Hybrid Symbolic-Neural Architectures*

Where domain-specific applications primarily narrow the problem space, a complementary line of work seeks to rebalance the architecture itself through hybrid symbolic-neural designs that explicitly integrate formal reasoning with neural language understanding. By coupling the semantic generalization capabilities of neural models with the precision and auditability of symbolic constraint systems, these architectures aim to improve both the correctness and traceability of generated test cases. They address a key failure mode of purely generative LLM approaches: the hallucination of requirement-inconsistent test steps.

Across the twenty-six primary studies surveyed, UMTG (Wang et al. [10]) is the only system to fully implement an integrated hybrid symbolic-neural architecture, with reported performance metrics in the NL-to-test domain. UMTG applies NLP-based semantic role labelling to extract control flow from use-case specifications, automatically generates Object Constraint Language (OCL) formal constraints from natural language sentences, and then applies constraint solving to produce executable test data. In two industrial automotive case studies, UMTG correctly translated 95 percent of use-case specification steps into formal OCL constraints and generated test cases that covered all expert-designed scenarios, as well as additional critical scenarios not previously considered [10]. This architectural pattern - neural NLP feeding a symbolic formal layer that drives test-data generation - represents the most tractable and empirically verified realization of hybrid symbolic-neural test-case generation reported in the literature.

Bougzime et al. [34] provide a broader architectural survey of neuro-symbolic AI paradigms, including retrieval-augmented generation, graph neural networks, and reinforcement learning, and discuss how symbolic reasoning can reduce hallucinations and improve determinism in generative AI systems. However, this work remains theoretical and architectural in nature: it does not implement a test case-generation system and reports no empirical performance metrics for that task. Other studies in this corpus that combine neural and rule-based elements do so primarily at the level of NLP pipelines; for example, Medeshetty et al. [17] compare rule-based information extraction with named-entity recognition rather than implementing full symbolic-neural integration.

Despite growing recognition of the importance of hybrid architectures, as evidenced by their identification as a key research gap in multiple surveys [22], [23], this area remains one of the least explored in the field. At present, UMTG is the only system providing rigorous empirical evidence that hybrid symbolic-neural architectures can successfully balance automation, correctness, and formal traceability in natural language-driven test case generation.

## VI. Comparative Analysis Of Tools and Frameworks

This section synthesizes the findings of Sections IV and V to answer RQ2 and RQ3 by comparing the tools and frameworks proposed for AI-based test case generation from natural language requirements. Building on the detailed discussion of techniques and architectures, the analysis contrasts representative approaches across eras - traditional NLP pipelines, machine learning systems, and LLM-based and hybrid solutions using a uniform set of evaluation dimensions. Rather than re-describing individual methods, this section focuses on cross-cutting distinctions in automation level, ambiguity handling, domain applicability, traceability, evaluation rigor, and trustworthiness, as summarized in the comparative tables that follow. This structured comparison exposes recurring trade-offs and limitations that motivate the research gaps identified in Section VII.

Over the past decades, the surveyed approaches have evolved through three distinct eras. The first, the Formalism Era, emphasized rule-based methods and formal specification techniques. This phase was followed by the Statistical/NLP Era, marked by a shift toward template-driven approaches and natural language processing pipelines. Most recently, the Hybrid/LLM Era has emerged, integrating symbolic reasoning, neural networks, and large language models.

Figure 6 shows the chronological distribution of the twenty-six confirmed primary studies across the three eras used in this survey. The empty Formalism Era band (2000–2012) reflects the effect of the strict I4 and I5 inclusion criteria rather than a lack of prior work: formal-specification approaches dominated that period but fall outside the present corpus because they do not jointly satisfy natural-language input and empirical test-generation validation. By contrast, the Statistical/NLP Era (2012–2022) and the Hybrid/LLM Era (2022–2025) contain the full set of confirmed primary studies, with the latter accounting for the most recent contributions.

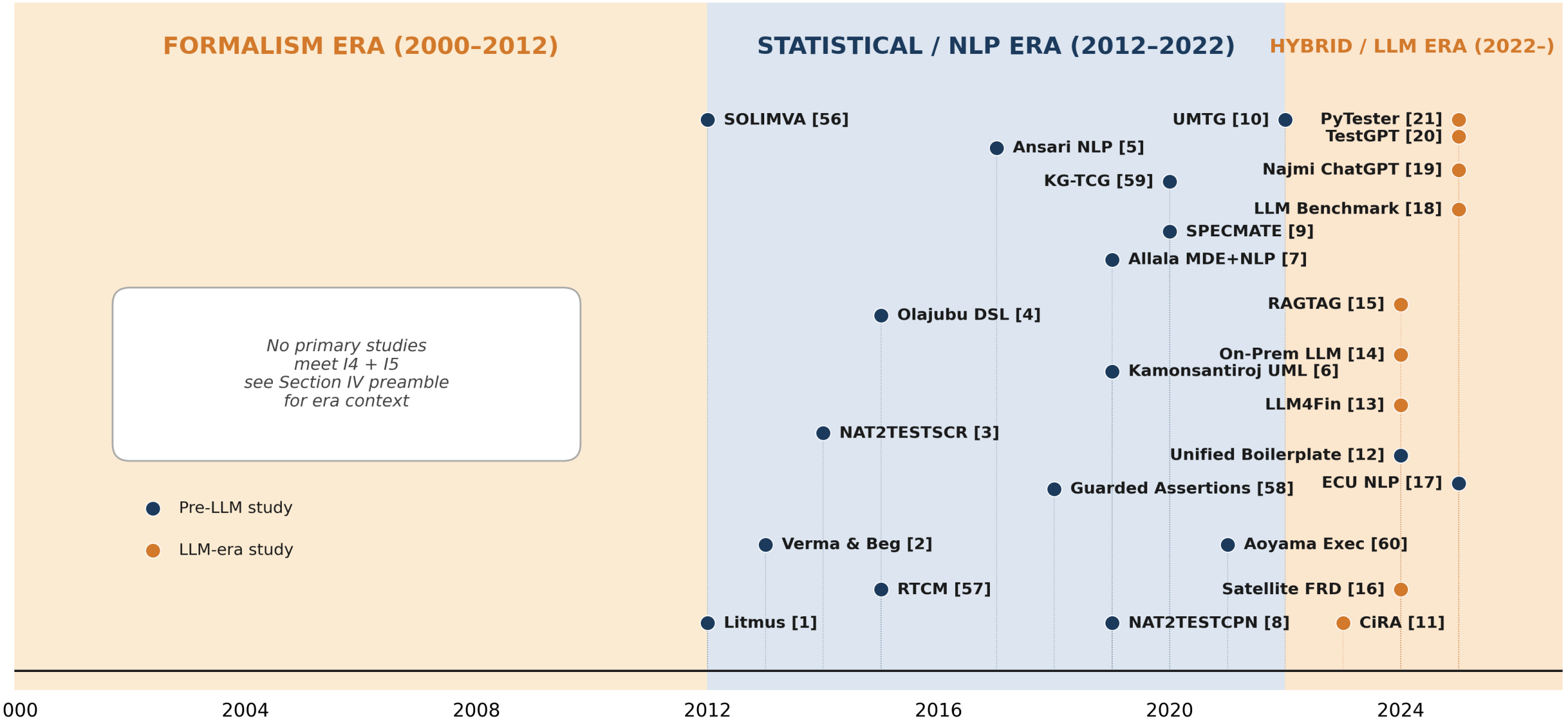

**Figure 6.** Chronological distribution of the twenty-six confirmed primary studies across the three eras (formalism, statistical/NLP, and Hybrid/LLM) used in this survey. The primary studies span the period 2012–2025.

The absence of data points in the Formalism Era reflects the scope of this survey: the I4 and I5 criteria exclude the formal-specification studies that dominated that period. However, their historical role is discussed in the background references [31], [32], [39], [41], [45], [47], [50], as well as the mapping studies [23], [27].

To ground this era-level evolution in concrete empirical evidence, Table II enumerates all 26 verified primary studies in chronological order. For each study, the table records the associated system or publication venue, the core NLP or AI technique employed, the type of input requirements, the generated test artifact and application domain, and the reported quantitative performance metrics. The rightmost column lists the reference number that anchors each study to the corresponding entry in the reference list.

TABLE II. TWENTY-SIX VERIFIED PRIMARY STUDIES - QUANTITATIVE DATA

| System / Venue | Technique | Input Type | Output / Domain | Verified Performance | Ref |
|---|---|---|---|---|---|
| Litmus (NLDB 2012) | 5-step NLP pipeline | NL functional requirements | Test cases / General | Testability → simplification → intent → synthesis → BVA. Qualitative. | 1 |
| SOLIMVA (SQJ 2012) | NL → Statecharts (combinatorial) | NL requirements | Test cases / Space (INPE) | NL → Statecharts via SOLIMVA; tests via GTSC. SWPDC satellite payload (QSEE). | 56 |
| Verma & Beg (IEEE ICETET 2013) | POS + parse tree | NL requirements | Test cases / General | POS + parse tree → knowledge graph → test inputs. | 2 |
| NAT2TESTSCR (SCP 2014) | CNL → SCR → T-VEC | SysReq-CNL requirements | Test vectors / Aerospace + Auto | Mutation 37.48% (TIS) → 95.52% (PC). 100% precision Embraer. | 3 |
| DSL from NL (ACM RACS 2015) | Domain-specific modelling | NL shall statements | Test cases / Aerospace (GE) | GE Aviation NL → DSL → tests via M2T. | 4 |
| RTCM (ACM ISSTA 2015) | Restricted NL + test modeling | Restricted NL (RTCM TCSs) | Test cases / Industrial (Cisco) | aToucan4Test: 9 TCSs → 246 executable test cases. Cisco Norway VCS. | 57 |
| NLP Keywords (IEEE AEEICB 2017) | Keyword-based NLP | NL functional SRS | Test cases / General | Keyword structure extraction from SRS. | 5 |
| Guarded Assertions (IEEE QRS 2018) | Guarded assertions (SAGA) | NL system-level requirements | Passive tests / Embedded (Bombardier) | NL → guarded assertions; temporal-constraint extraction. Bombardier trains. | 58 |
| UML Memorization (ACM ICGDA 2019) | Memorization + UML activity diagrams | UML + NL annotations | Test cases / Concurrent systems | Memorization approach for UML activity diagram test generation. | 6 |
| MDE + NLP (IEEE COMPSAC 2019) | MDE + Stanford CoreNLP | NL user requirements | Test cases / General | NL reqs → metamodel → test cases via M2M transformation. | 7 |
| NAT2TESTCPN (SCP 2019) | CNL → Coloured Petri Net | SysReq-CNL requirements | Test cases / Aerospace + Auto | ~80% VM/NPP/PC, ~54% TIS, 85.07% peak PC. | 8 |
| SPECMATE (IEEE ICST 2020) | NLP pattern + CEG | NL acceptance criteria | Test cases / Industrial (Allianz) | 56% of 604 TCs. Missing negative TCs added. 961 user stories. | 9 |
| KG-TCG (ACM CoDS 2020) | Knowledge graph + NER | NL (domain) requirements | Test cases / Automotive (Bosch) | KG intermediate from requirements; handles missing info. Bosch automotive. | 59 |
| Executable TCG (IEEE CCNC 2021) | EBNF → propositional net → decision table | NL specifications | Executable tests / HVAC (SPLE) | NL spec → decision-table model → executable tests + execution env. HVAC SPLE. | 60 |
| UMTG (IEEE TSE 2022) | NLP SRL → OCL + solver | NL use-case (RUCM) | Tests + traceability / Automotive | 95% of use-case steps → OCL constraints. All expert + critical scenarios. | 10 |
| CiRA (JSS 2023) | RoBERTa conditional | Informal NL requirements | Acceptance tests / Industrial | 71.8% of 578 TCs. 80 missed TCs found. 3 partners. | 11 |
| Unified Boilerplate (JSS 2024) | Rupp + EARS boilerplate | NL requirements (PURE) | Test templates / General | Pointis: 61.7%, Mdot: 50.0%, Npac: 10%. | 12 |
| LLM4Fin (ICSE Companion 2024) | Domain-tuned LLM + KB | NL financial rules | Acceptance tests / FinTech | Avg BSC 91.89%; peak 98.18% (D3). ChatGPT avg 49.65%. 20 min → 7 sec. | 13 |
| On-Prem LLM (IEEE ICIIBMS 2024) | LLaMA-3 70B + prompt chain | NL user stories | Secure tests / Enterprise | 79.6% Llama-3-70B (Tbl. II); 90.3% Prompt-Chain+2-shot (Tbl. III). | 14 |

| System / Venue | Technique | Input Type | Output / Domain | Verified Performance | Ref |
|---|---|---|---|---|---|
| RAGTAG (IEEE RE 2024) | RAG + LLMs (GPT-3.5/4.0) | NL requirements (bilingual) | Test scenarios / Industrial (Austrian Post) | Expert: relevant/feasible. BLEU=0.092, ROUGE=0.419, METEOR=0.516. | 15 |
| Satellite FRD (IEEE ICECCME 2024) | NLP (NER+POS+DP) + Few-Shot+CoT LLM | Satellite FRD (NL) | Test cases / Space (ISRO) | GPT-4=90%. Gemini-Pro=60%. Llama-2-70B=40%. Mistral-7B=20%. | 16 |
| ECU NLP (IEEE ICHMS 2025) | Rule-based IE + NER SVM | NL ECU feature elements | Test specs / Automotive | 95%/75%/60% by complexity. SVM 77.3% (p=0.037). 24 hr → 5 min. | 17 |
| LLM Benchmark (ISEC 2025) | 6-LLM comparison | 25 NL requirements | Coverage metrics / Multi-domain | Gemini decision 85.2%, cond. 82.9%. All LLMs MCDC ≤ 50.2%. | 18 |
| ChatGPT Framework (JCS 2025) | ChatGPT + NLP + DL | NL requirements | Test cases / OCR | 4-stage ChatGPT+NLP+DL pipeline. Vol 21(5) 2025. | 19 |
| TestGPT (JSS 2025) | Fine-tuned GPT-3.5 | NL requirements | Unit tests / General | 78.5% syntactic, 67.09% alignment, 61.7% coverage, 18.9% mutation. | 20 |
| PyTester (JSS 2025) | Deep reinforcement learning | Text descriptions | Test cases / General | RL text-to-testcase. outperforms GPT-3.5 on APPS benchmark. | 21 |

*Abbreviations used in Tables II–IV: NL = natural language; reqs = requirements; TC(s) = test case(s); CNL = controlled natural language; SCR = Software Cost Reduction notation; OCL = Object Constraint Language; RAG = retrieval-augmented generation; M2M = model-to-model; FRD = functional requirements document; BSC = business-scenario coverage; MCDC = modified condition/decision coverage; LLM = large language model.*

b) Analysis of Tools & Frameworks - Era Evolution: Table II presents the 26 verified primary studies in chronological order, showing for each study the corresponding system or publication venue, the principal NLP or AI technique, the input requirements type, the generated test artifact and application domain, and the reported quantitative performance metrics.

## A. *Analysis of Table II - What the Twenty-One Primary Studies Collectively Show*

Three patterns emerge from the chronological quantitative data in Table II. First, the field's center of gravity has decisively shifted into the Hybrid/LLM Era. Of the twenty-six primary studies, eleven pre-date 2020; the remaining fifteen appear between 2020 and 2025, and nine of those fifteen are LLM-based or LLM-hybrid approaches (Xue et al.'s LLM4Fin [13], Yin et al.'s on-premises LLaMA-3 70B [14], Arora et al.'s RAGTAG [15], Shakthi et al.'s satellite-FRD study [16], Korraprolu et al.'s six-LLM benchmark [18], Najmi and El-Dosuky's ChatGPT framework [19], Alagarsamy et al.'s TestGPT [20], Takerngsaksiri et al.'s PyTester [21], and Fischbach et al.'s CiRA [11]). The absolute peak numbers reported 98.18 percent peak Business-Scenario Coverage by LLM4Fin on Dataset 3, 95 percent OCL-translation accuracy by UMTG [10], 95.52 percent peak mutation score by NAT2TESTSCR [3] on the Embraer example, and 90.3 percent coverage by the on-premises LLaMA-3 70B system. They all cluster above 90 percent, suggesting that the ceiling for what NL-driven TCG can achieve lies not in the technique itself but in the measurement regime and the domain.

Second, peak numbers and averages diverge sharply in LLM studies and converge in formal, controlled-NL studies. LLM4Fin's 98.18 percent peak versus 91.89 percent average (Dataset 3 vs. five-dataset average, [13]) and the GPT-4 90 percent versus Mistral-7B 20 percent spread on identical satellite FRDs [16] are representative: LLM outputs are more variable than formal CNL outputs and averaging across datasets or models erodes the peak. The deterministic-CNL approaches, NAT2TESTSCR [3] and NAT2TESTCPN [8], show the opposite pattern: their mutation scores cluster tightly around 80 percent across four case studies, with a single outlier at 95.52 percent. This divergence is the empirical signature of the trade-off between traceability and expressiveness, as analyzed quantitatively in section VI-B.

Third, industrial and safety-critical domains dominate the 2020-2025 cohort: Allianz Deutschland (SPECMATE [9], CiRA [11]), Austrian Post (RAGTAG [15]), Ericsson and Leopold Kostal (CiRA [11]), automotive ECUs (Medeshetty et al. [17], UMTG [10]), FinTech compliance (LLM4Fin [13]), and ISRO satellite FRDs (Shakthi et al. [16]). This concentration of industrial case studies is a strength of the corpus: validation is no longer limited to academic toy examples. But it also exposes the absence of published reproducibility packages. Most industrial studies report aggregate metrics without releasing the underlying requirements or test case data, which is a structural obstacle to the complexity-stratified benchmark that R3 proposes in section VIII-C.

## B. *Related Work Summary*

While the preceding analysis focuses on the twenty-six confirmed primary studies, it is also necessary to situate these contributions within the broader research landscape. This subsection, therefore, summarizes representative related and background work that informs, contextualizes, or bounds the scope of AI-based test case generation from natural-language requirements.

Table III cross-references the twelve most representative related works, including selected primary studies already tabulated in Table II, as well as adjacent surveys and background references, using a seven-column schema that captures approach, input, output, evaluation design, core assumption, principal limitation, and inclusion status. The table is original to this survey (Contribution C2) and serves as the mechanism for comparing primary studies, related work, and background sources using uniform criteria.

TABLE III. RELATED WORK SUMMARY - TWELVE REPRESENTATIVE STUDIES (CONTRIBUTION C2)

| Ref | Approach | Input | Output | Evaluation | Key Assumption | Key Limitation | Status |
|---|---|---|---|---|---|---|---|
| [10] | UMTG: NLP SRL → OCL | NL use-case specs | Tests + traceability | 2 industrial studies | RUCM template required | Domain model required | PRIMARY |
| [3] | NAT2TESTSCR: CNL → SCR | CNL requirements | Test vectors | 4 case studies; mutation | Engineers write CNL | CNL training needed | PRIMARY |
| [11] | CiRA: RoBERTa conditional | Informal NL requirements | Acceptance tests | Industrial: 3 partners | Conditional structure present | Conditional-centric only | PRIMARY |
| [9] | SPECMATE: NLP + CEG | NL acceptance criteria | Test cases | 604 TCs; 56% | User stories with ACs | AC-type coverage | PRIMARY |
| [15] | RAGTAG: RAG + LLMs | NL (bilingual) | Test scenarios | 4 experts; 5 dimensions | Domain context retrievable | Action-sequence accuracy | PRIMARY |
| [13] | LLM4Fin: domain LLM | NL financial rules | Acceptance tests | BSC on 5 datasets | Domain knowledge pre-coded | FinTech only | PRIMARY |
| [14] | LLaMA-3 70B on-prem | NL user stories | Secure test cases | Coverage up to 90.3% | GPU VRAM required | Human eval only | PRIMARY |
| [22] | Yang RBTG Survey | N/A (survey) | Taxonomy | SLR; 267 papers | Wide scope | LLM hallucination unaddressed | RELATED WORK |
| [24] | Boukhlif NLP-testing SLR | N/A (survey) | Classification | PRISMA; 24 papers | NLP testing feasible | Pre-LLM scope | RELATED WORK |
| [27] | Garousi NLP-testing mapping | N/A (mapping) | 67-paper map | IST Q1, 2020 | NLP applied to testing | Broader than TCG-only | RELATED WORK |
| [49] | Cajica mutation TCG | Formal specs | Test cases | Coverage metrics | Formal spec available | Fails I4 + I5 | BACKGROUN D |
| [50] | Helke Z + Isabelle | Z formal specification | Test cases | Small experiment | Z spec available | Fails I4 + I5 | BACKGROUN D |

c) Related Work Summary: Table III compares twelve key related works using a seven-column framework covering methodology, inputs, outputs, evaluation, assumptions, limitations, and inclusion status. Created as Contribution C2 of this survey, it provides a uniform basis for comparing primary studies, related surveys, and background references.

The twelve rows in Table III span three inclusion categories: PRIMARY (seven entries), RELATED WORK (three entries), and BACKGROUND (two entries). Read vertically, the table exposes two structural observations that cut across these categories.

First, the "Key Limitation" column converges on a small number of recurring weaknesses. Formal-specification-only background work consistently fails the natural language and empirical test generation criteria (Fails I4 + I5), as exemplified by Cajica [49] and Helke [50]. Older surveys, such as Boukhlif [24] and Garousi [27], are characterized by a pre-LLM scope or by coverage broader than test case generation alone, limiting their ability to address LLM-specific challenges. In contrast, LLM-era primary studies exhibit domain-restriction limitations, including FinTech-only applicability in LLM4Fin [13] and enterprise-constrained deployment assumptions in the on-premises LLaMA-3 system [14]. This clustering of limitations constitutes cross-cutting evidence motivating R3's call for a multi-domain, complexity-stratified benchmark.

Second, the "Key Assumption" column reveals the implicit contract each approach requires users to accept to make requirements-to-test-case generation tractable. These assumptions include Controlled Natural Language authoring in [3], RUCM templating in [10], explicit conditional structure in [11], retrievable domain context in [15], pre-coded domain knowledge in [13], and GPU VRAM availability in [14]. Collectively, these assumptions demonstrate that the primary studies have internalized a fundamental lesson from the background formal-methods literature: unrestricted natural language is too expressive to map cleanly to executable tests without additional constraints. Each approach, therefore, imposes explicit structural, domain, or infrastructural assumptions to regain tractability.

The positioning of this survey between related-work surveys and confirmed primary studies is therefore not incidental. The primary studies inherit formal-method rigor from the background layer while simultaneously extending the broad natural language applicability characterized in LLM-era surveys. Table III makes this inheritance explicit and provides the interpretive bridge between foundational work and the empirical, LLM-driven systems evaluated in this paper.

## C. Six-Criteria Comparison and Gap Analysis

The comparative patterns and recurring limitations identified across the primary studies and related work motivate the need for a uniform evaluative framework. To make these trade-offs explicit and systematically comparable across techniques and eras, we define a set of six evaluation criteria that capture the core dimensions along which existing approaches differ.

The six criteria are defined to be consistently applicable across the majority of primary studies and directly tied to the research questions: K1 automation level (manual | semi auto | full); K2 ambiguity handling (Low | Medium | High); K3 domain applicability (Narrow | Medium | Broad); K4 traceability (Low | Medium | High); K5 evaluation thoroughness (Low | Medium | High); and K6 hallucination control (Low | Medium | High).

Criteria that did not apply across $\geqslant 80$ percent of studies consistently were removed or merged.

Table IV applies the six criteria uniformly to eight representative primary studies, ranging from deterministic controlled natural language pipelines (NAT2TESTSCR), through conditional parse NLP approaches (CiRA), retrieval augmented LLM systems (RAGTAG, LLM4Fin), on-premises deployments (Yin et al.), multi-LLM benchmarks (Korraprolu et al.), to fine-tuned LLM solutions (TestGPT). The colored bottom row summarizes the overall pattern across criteria; the detailed gap analysis immediately following the table explains the structural trade-offs that the summary row only hints at.

The eight studies in Table IV were not chosen arbitrarily; they were selected by purposive maximum-variation sampling from the twenty-six primary studies, using four explicit criteria. First, era coverage: the subset spans all three evolutionary eras, from the deterministic Statistical/NLP-era pipelines (NAT2TESTSCR [3], UMTG [10]) to the most recent Hybrid/LLM-era systems (RAGTAG [15], LLM4Fin [13], the on-premises LLaMA-3 system [14], the multi-LLM benchmark [18], and TestGPT [20]). Second, architectural diversity: the subset deliberately includes one instance of each major architectural class identified in Sections IV and V - controlled natural language translation, model-based constraint solving, conditional-parse NLP, retrieval augmented generation, domain-tuned LLM, on-premises quantized LLM, multi-model benchmark, and fine-tuned LLM - so that the comparison captures the full design space rather than one region of it. Third, evaluation sufficiency: only studies that reported sufficient methodological and empirical detail to be coded on all six criteria, without 'not reported' cells, were included, thereby keeping the matrix fully populated and therefore interpretable. Fourth, influence and industrial validation: each selected study is either a frequently cited reference point or reports a real industrial or safety-critical deployment. This is a deliberate analytical design choice rather than a convenience sample: the six criteria require interpretive High/Medium/Low coding of each study, and several of the remaining thirteen primary studies do not report sufficient detail to be coded on all six dimensions without introducing 'not reported' cells that would weaken rather than strengthen the comparison.

TABLE IV. SIX-CRITERIA COMPARISON - EIGHT REPRESENTATIVE PRIMARY STUDIES (CONTRIBUTION C3)

| Ref | Technique | K1 Automation | K2 Ambiguity | K3 Domain | K4 Traceability | K5 Evaluation | K6 Hallucination |
|---|---|---|---|---|---|---|---|
| 10 | UMTG: NLP SRL → OCL | Semi-auto | Medium | Medium - 2 studies | HIGH - OCL; RTTM artifact | HIGH - industrial | HIGH - formal layer |
| 3 | CNL + SCR + T-VEC | Full | HIGH - CNL | Low - CNL required | HIGH - SCR tables | HIGH - mutation; 4 domains | HIGH - deterministic |
| 11 | CiRA RoBERTa | Full | HIGH - explicit condition parse | Medium - 3 industry partners | Medium - requirements linked to TCs | HIGH - 578 TCs; 3 partners | Medium - NLP rules + RoBERTa |
| 15 | RAGTAG RAG+LLM | Full | Medium - RAG retrieval helps | Medium - bilingual industrial | Low - no RTTM produced | HIGH - 4 experts; 5 dims | Medium - RAG anchors context |
| 13 | LLM4Fin domain-tuned | Full | Medium - domain knowledge helps | Low - FinTech only | Low - none reported | HIGH - BSC; 5 datasets | LOW - LLM; uncontrolled |
| 14 | LLaMA-3 70B on-prem | Full | Low - no explicit mechanism | Low - enterprise only | Low - none reported | Medium - 90.3%; human eval | LOW - LLM; no control |
| 18 | LLM benchmark - 6 models | Full | Low - no mechanism | Medium - multi-domain | Low - none reported | HIGH - 4 metrics; 25 reqs | LOW - correctness not evaluated |
| 20 | TestGPT fine-tuned | Full | Medium - feedback loop helps | Medium - needs source code | Low - none reported | Medium - syntax + coverage | LOW - LLM; uncontrolled |

In summary, none of the eight studies attains a high rating on all six criteria. Traceability (K4) is low in five studies and medium in one, and hallucination control (K6) is low in four studies and medium in two, making them the weakest dimensions in the comparison. CiRA [11] and RAGTAG [15] achieve high evaluation thoroughness with medium hallucination control by constraining generation through explicit NLP structure and retrieval grounding,

| Ref | Technique | K1 Automation | K2 Ambiguity | K3 Domain | K4 Traceability | K5 Evaluation | K6 Hallucination |
|---|---|---|---|---|---|---|---|

respectively, but neither provides formal traceability. Overall, the comparison indicates a consistent trade-off: as input expressiveness increases, formal traceability decreases.

Although the six-criteria matrix is applied to eight representative studies for readability, its central patterns are consistent with the full twenty-six-study corpus characterized in Table II, and the matrix is therefore not a generalization from an unrepresentative sample. The traceability deficit is corpus-wide: across all twenty-six studies, only the controlled-input and model-based approaches (NAT2TESTSCR [3], NAT2TESTCPN [8], UMTG [10]) report a formal requirements-to-test traceability artifact, while every fully generative LLM-based study omits one. The absence of explicit hallucination handling is likewise corpus-wide: no primary study outside the external-anchoring group (UMTG [10], NAT2TESTSCR [3], CiRA [11], RAGTAG [15]) reports a hallucination detection or mitigation mechanism. Finally, the accuracy-degradation pattern in Gap finding 3 is confirmed beyond the subset by Medeshetty et al. [17] and Shakthi et al. [16], which are part of the twenty-six-study corpus but not of the eight-study matrix. The eight-study table is thus an expository distillation of the full corpus, and its conclusions hold at corpus scale.

d) Six Criteria Comparison: In Table IV, we defined six evaluation criteria that capture the core dimensions along which existing approaches differ. These criteria were applied to eight purposely selected and representative studies within the twenty-six primary studies set.

### *D. Detailed Gap-Finding Analysis of Table IV*

Building on the six-criterion comparison introduced above, the matrix in Table IV exposes structural gaps that no single primary study fully addresses. Each gap identified below is diagnosed with reference to specific cell values in the table and maps directly to one of the four research gaps (G1-G4) formalized in section VII.

- Gap finding 1 - the automation vs traceability diagonal.

Reading the K1 (Automation) and K4 (Traceability) columns jointly yields the clearest structural pattern in the matrix. NAT2TESTSCR [3] achieves Full automation (K1) and High traceability (K4), but only by restricting inputs to Controlled Natural Language, resulting in Low domain applicability (K3). UMTG [10] similarly achieves High traceability with semi-automated processes by relying on RUCM-templated use cases. In contrast, every study that relaxes input restrictions to unconstrained natural language - LLM4Fin [13], the on-premises LLaMA 3 system [14], Korraprolu et al.'s multi-LLM benchmark [18], and TestGPT [20] exhibits Low traceability (K4). The resulting pattern is effectively diagonal: as approaches move along the automation and expressiveness axis from [3] to [18], formal traceability is progressively traded for greater input flexibility. No row in the matrix escapes this trade-off. This diagonal provides the direct empirical evidence for Gap G2 (VII) and motivates the RTTM pipeline recommendation R2 (VIII B).

- Gap finding 2 - hallucination control collapses in pure-LLM approaches.

Column K6 (Hallucination control) shows Low ratings for four of the eight studies: LLM4Fin [13], the on-premises LLaMA-3 system [14], Korraprolu et al.'s multi-LLM benchmark [18], and TestGPT [20], all pure-LLM systems. In contrast, the systems achieving High or Medium hallucination control - UMTG [10], NAT2TESTSCR [3], CiRA [11], and RAGTAG [15] do so by introducing an explicit external anchoring mechanism. UMTG relies on OCL constraints as a downstream verifier; NAT2TESTSCR uses SCR tables; CiRA applies explicit cause-effect graph parsing; and RAGTAG employs retrieval augmented generation to bind outputs to project-specific context. While the mechanisms differ, the pattern is consistent: effective hallucination control requires architectural commitments beyond the LLM itself. The combined HaluEval finding of a 19.5 percent general-response hallucination rate [53] and the 90-percent versus 20-percent performance spread between GPT-4 and Mistral-7B on identical satellite FRDs reported by Shakthi et al. [16] establishes the external quantitative baseline against which any future TCG-specific hallucination benchmark - R1 (in VIII-A) must be calibrated.

- Gap finding 3 - evaluation thoroughness and reproducibility are inversely related.

At first glance, column K5 (Evaluation thoroughness) suggests a mature field: six of the eight studies score High. However, the studies achieving High K5 through industrial case study evaluations - UMTG's automotive studies [10], CiRA's three partner deployment [11], RAGTAG's Austrian Post study [15], and LLM4Fin's five-dataset FinTech evaluation [13] are precisely those whose underlying requirements and test case data cannot be released for reproducibility. By contrast, the two studies with publicly available benchmarks, Korraprolu et al.'s twenty-five requirement multi-LLM benchmark [18] and TestGPT [20], achieve High and Medium K5 scores, respectively, but trade off industrial case-study depth. This constitutes the reproducibility vs industrial depth trade-off, providing concrete evidence for Gap G3 (accuracy degrades with requirement complexity) and motivating the complexity-stratified benchmark recommendation R3 (VIII C). The benchmark proposed in R3 is explicitly designed to occupy the intersection of multi-domain scope, public availability, and complexity stratification that no existing study currently achieves.

- Gap finding 4 - compliance, on-premises deployment, and the enterprise ceiling.

Within Table IV, only Yin et al.'s on-premises LLaMA-3 system [14] directly addresses the combined constraints of data residency, export-control sensitivity, and deployable model size imposed by real enterprise environments. This evidence is reinforced beyond the table by Shakthi et al.'s ISRO satellite-FRD deployment [16], discussed in Section V, which is part of the twenty-six-study corpus but is not one of the eight studies tabulated here. The on-premises LLaMA-3 system scores Low on K3 (domain applicability) because it does not generalize beyond its specific deployment context, and Low on K4 (traceability) because it reports no traceability artifacts. Although this system achieves 90.3 percent coverage, the highest empirically demonstrated ceiling in Table IV for a deployable enterprise-grade configuration, the absence of reported regulatory-compliance evidence (e.g., GDPR, HIPAA, DO-178C) across all studies constitutes the structural signal for Gap G4 and motivates the regulation-compliant deployment-blueprint recommendation R4 (VIII-D).

Taken together, the four gap findings- the automation-vs-traceability trade-off, the collapse of hallucination control without external anchoring, the reproducibility-vs-industrial-depth tension, and the compliance void reveal a common architectural limitation. The field has not yet produced an approach that is simultaneously highly automated, formally traceable, domain-broad, and regulation-compliant. Section VII builds directly on this comparative evidence by formalizing the four corresponding research gaps, G1-G4, and grounding each in the primary-study corpus.

Given that the foregoing discussion draws corpus-level inferences from the patterns observed in Table IV, a methodological clarification is warranted. The six-criterion comparison reported in Table IV is applied to a purposively selected subset of eight studies rather than to the full set of twenty-six primary studies. This constitutes a deliberate analytical choice because the six criteria require fine-grained interpretive coding, limiting the matrix to studies that can be coded completely on all criteria and avoiding the proliferation of "not reported" cells that a compulsory full-corpus matrix would introduce. The potential validity threat that subset selection might distort the gap analysis is mitigated in two respects. First, the subset was constructed through maximum-

## VII. Research Challenges, Gaps, and Evidence-Based Analysis

The comparative analysis in Section VI shows that, despite substantial progress in NLP and LLM-based test generation, several structural limitations remain unresolved. Existing approaches continue to trade automation for traceability, broad applicability for architectural control, and industrial realism for reproducibility.

Building on that evidence, this section addresses RQ4 by formalizing four research gaps: G1 (hallucination), G2 (traceability), G3 (complexity sensitivity), and G4 (compliance). Together, these gaps capture the principal obstacles to trustworthy, scalable, and deployable AI-based test case generation from natural language requirements.

Each gap is grounded in specific quantitative and architectural evidence from the primary studies and, in turn, motivates the actionable recommendations developed in Section VIII.

### A. *Gap G1 - Absence of Hallucination Detection in LLM-Generated Test Cases*

Gap G1 concerns the absence of explicit hallucination-detection mechanisms in LLM-based test case generation. The six-criterion analysis shows that hallucination control remains weakly addressed: in Table IV, four of the eight representative studies - LLM4Fin [13], the on-premises LLaMA-3 system [14], Korraprolu et al.'s multi-LLM benchmark [18], and TestGPT [20] - receive Low ratings on criterion K6 (hallucination control). These systems primarily rely on unconstrained generative inference and provide no explicit architectural mechanism for detecting or mitigating requirement-inconsistent test steps.

In contrast, the approaches that achieve High or Medium hallucination control - UMTG [10], NAT2TESTSCR [3], CiRA [11], and RAGTAG [15] all introduce an external anchoring mechanism. UMTG employs OCL constraints as a downstream verifier; NAT2TESTSCR relies on SCR tables; CiRA uses explicit cause-effect graph parsing; and RAGTAG applies retrieval augmented generation to bind outputs to project-specific requirement context. Although these mechanisms differ in implementation, the pattern is consistent: effective hallucination control requires architectural commitments beyond the LLM itself.

This gap is further contextualized by external empirical evidence. Li et al.'s HaluEval benchmark reports that 19.5 percent of general LLM responses (977 of 5,000 evaluated responses, verbatim) contain hallucinations [53], underscoring the prevalence of the problem. In the domain of test case generation, Shakthi et al. [16] demonstrate a 90 percent versus 20 percent correctness gap between GPT 4 and Mistral 7B on identical satellite Functional Requirement Documents, confirming that hallucination frequency is strongly model-dependent and amplified by input complexity.

Despite this evidence, no primary study provides a test-case-specific hallucination benchmark or a systematic post-hoc detection mechanism tailored to NL-to-TCG pipelines. This absence constitutes Gap G1 and motivates Recommendation R1, which calls for the development of a dedicated hallucination evaluation benchmark for LLM-generated test cases (VIII A).

### B. *Gap G2 - No System Achieves Both Full Automation and Formal Traceability*

Gap G2 concerns the absence of any approach that simultaneously achieves full automation and formal traceability for unconstrained natural-language requirements. UMTG [10] and NAT2TESTSCR [3] provide formal requirements-to-test traceability, but only by imposing controlled or templated input conditions. CiRA [11] preserves a partial link between extracted conditional structures and generated acceptance tests and therefore represents the closest approximation to traceability among NLP-based systems operating on unconstrained natural language; however, it does not produce a formal requirements-to-test traceability matrix.

In contrast, all fully generative LLM-based systems produce no traceability artifact [14], [22], reflecting the absence of an auditable linkage between input requirements and generated test cases. This loss of traceability constitutes a systematic limitation of current end-to-end generative approaches.

External architectural precedents demonstrate that traceability is achievable but remains decoupled from test case generation. Sawada et al. [30] demonstrate operational NLP + ML-based requirements to test traceability at NASA JPL, providing a concrete example of scalable traceability support in an industrial setting. More recently, Ge et al. [55] (IEEE TSE 2025) demonstrate LLM-based cross-level requirements tracing, establishing an architectural precedent for integrating traceability mechanisms within the LLM paradigm. However, neither approach couples' traceability directly with natural-language-driven test case generation.

Taken together, these findings show that no existing approach simultaneously achieves full automation and formal traceability when operating on unconstrained natural language requirements. This absence constitutes Gap G2 and directly motivates Recommendation R2, which calls for an auditable requirements-to-test traceability matrix (RTTM) pipeline (VIII-B).

### C. Gap G3 - Accuracy Degrades with Requirement Complexity

Gap G3 concerns the systematic degradation of test generation accuracy as requirement complexity increases. Medeshetty et al. [17] provide the clearest direct quantification of this effect, reporting a stepwise decline from 95 percent to 75 percent to 60 percent accuracy across single-signal, up-to-four-signal, and more-than-four-signal requirement categories. This result establishes a direct empirical relationship between requirement complexity and test generation accuracy.

Evidence from LLM-based approaches reinforces this pattern. Korraprolu et al. [18] confirm that all six general-purpose LLMs evaluated fail to exceed 50.2 percent Modified Condition/Decision Coverage (MCDC) for complex, multi-condition requirements, with Gemini achieving the highest reported score. Similarly, Shakhti et al. [16] report a pronounced 90 percent to 20 percent drop in correctness between GPT 4 and Mistral 7B when evaluated on ten identical satellite Functional Requirement Documents, demonstrating that model capability and input complexity interact nonlinearly in practice.

Despite the consistency of these findings, empirical evaluation remains constrained by limitations in the dataset. Publicly available requirements datasets, including PURE [54] and PROMISE, are small-scale and limited in domain coverage, restricting their ability to support systematic, cross-complexity evaluation [26], [23]. As a result, accuracy degradation with increasing requirement complexity is well documented but insufficiently benchmarked across domains and model classes.

Taken together, these observations show that no existing study provides a standardized, complexity-stratified evaluation framework for NL-to-TCG systems. This absence constitutes Gap G3 and directly motivates Recommendation R3, which calls for the development of a shared, multi-domain benchmark with explicit stratification by requirement complexity (VIII C).

### D. Gap G4 - Ethical, Security, and Regulatory Compliance

Gap G4 concerns the absence of regulation-compliant, enterprise-ready deployment patterns for AI-based test case generation. Within Table IV, only Yin et al.'s on-premises LLaMA-3 system [14] directly addresses the combined constraints of data residency, export-control sensitivity, and deployable model size that real enterprise environments impose. This evidence is reinforced beyond the table by Shakhti et al.'s ISRO satellite-FRD deployment [16], which is discussed in Section V as part of the twenty-six-study corpus but is not one of the eight tabulated studies. The on-premises LLaMA-3 system scores Low on K3 (domain applicability) because it does not generalize beyond its specific deployment context, and it is silent on K4 (traceability) because it produces no requirements-to-tests traceability matrix.

Despite these limitations, the on-premises LLaMA 3 system reported by Yin et al. [14] achieves 90.3 percent coverage, representing the highest empirically demonstrated ceiling in Table IV for a deployable enterprise-grade configuration. This result establishes an upper bound on performance under realistic enterprise constraints but also highlights the narrow applicability and limited auditability of current compliant deployments.

More broadly, cloud-based LLM solutions introduce unresolved risks related to data leakage, licensing, bias, and regulatory exposure. No study in Table IV reports concrete evidence of compliance with regulatory standards such as GDPR, HIPAA, or DO 178C, underscoring the absence of validation frameworks for regulated deployment contexts. This lack of reported compliance evidence constitutes a structural barrier to adoption in safety-critical and regulated domains.

Taken together, these observations show that no existing approach satisfies enterprise-grade requirements for security, compliance, and auditability while maintaining broad domain applicability and traceability. This absence constitutes Gap G4 and directly motivates Recommendation R4, which calls for the development of regulation-compliant, on-premises deployment blueprints for AI-based test-case generation (VIII D).

## VIII. Guidelines, Recommendations, and Emerging Trends

This section addresses RQ4, which concerns how the research gaps identified in the preceding analysis should be addressed. The four evidence-based research gaps formalized in Section VII motivate a corresponding set of actionable guidelines, each designed to directly address a specific structural limitation observed in current approaches to AI-based test-case generation from natural-language requirements.

Figure 7 presents a one-to-one mapping between the four research gaps (G1-G4) and the recommended guidelines (R1-R4), making the correspondence between identified limitations and proposed remedies explicit. Gap G1 highlights the absence of hallucination detection in LLM-generated test cases and motivates R1: the development of HalluTCG-Bench, a test-case-generation-specific hallucination benchmark. Gap G2 captures the structural tension between full automation and formal traceability, leading to R2, an auditable requirements-to-test traceability matrix (RTTM) pipeline. Gap G3 identifies systematic accuracy degradation as requirement complexity increases, motivating R3, a complexity-stratified benchmark spanning multiple domains. Finally, Gap G4 exposes unresolved ethical, security, and regulatory constraints, which R4 addresses through regulation-compliant, on-premises deployment blueprints.

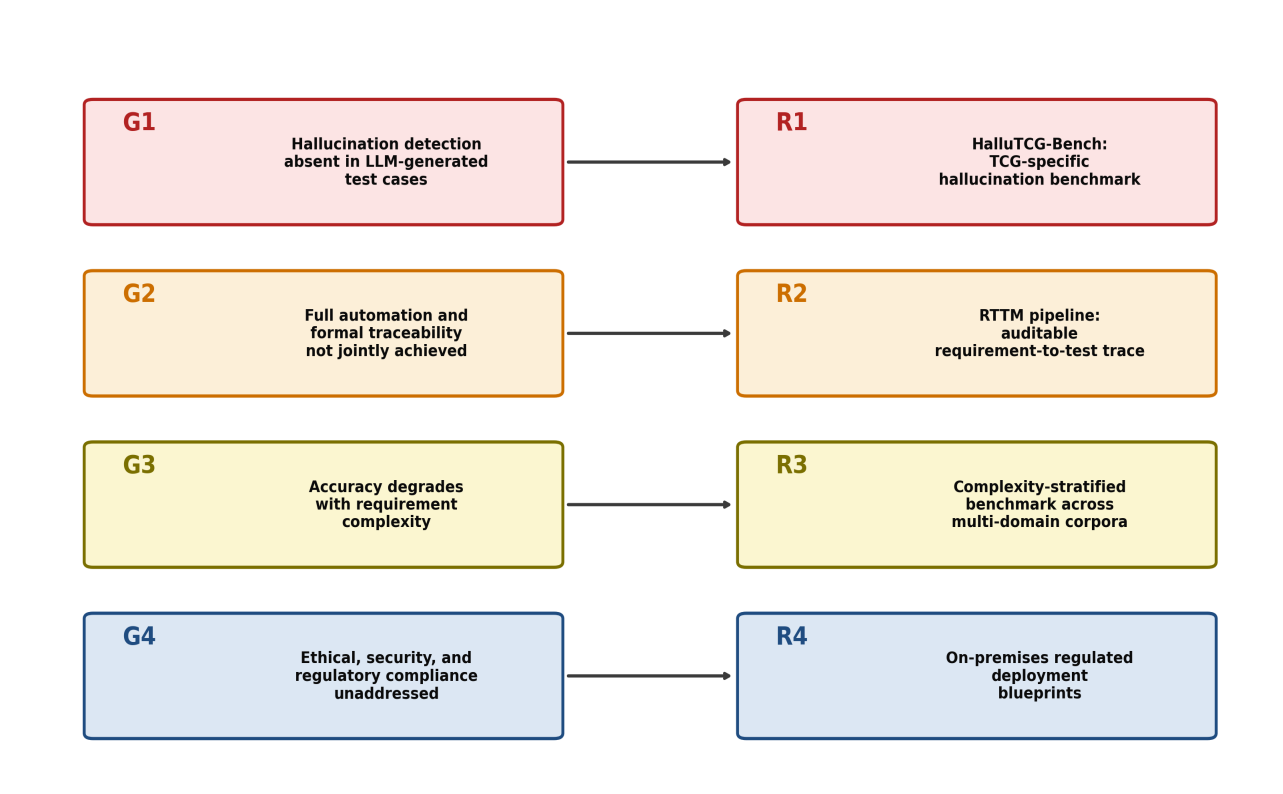

Figure 7. Mapping of the four identified research gaps (G1–G4) to the corresponding recommendations (R1–R4) and deliverable artifacts.

Each recommendation is defined as a concrete, implementable deliverable artifact whose construction would directly address its associated research gap and advance the state of practice toward more reliable, traceable, and deployable AI-based test case generation.

### *A. R1 - Develop HalluTCG-Bench (addresses G1)*

This recommendation proposes the development of a test-case-generation-specific benchmark for evaluating hallucination based on real-world natural-language requirements across multiple application domains. Empirical evidence from existing studies establishes a clear baseline for such a benchmark. The retrieval-augmented mitigation evaluated by Arora et al. [15] and the model capability variation reported by Shakthi et al. [16] together demonstrate that hallucination behavior is both architecture-dependent and domain-dependent. In particular, the 90 percent versus 20 percent correctness gap between GPT-4 and Mistral-7B on identical satellite Functional Requirement Documents confirms that hallucination rates cannot be treated as a model-agnostic phenomenon.

Li et al.'s HaluEval benchmark [53] provides a validated annotation methodology that can be adapted to the NL-to-TCG context, offering a foundation for systematically labeling hallucination-related errors. Complementing this, PyTester [21] demonstrates a reinforcement-learning-based optimization pathway that can support controlled experimentation with hallucination mitigation strategies once such a benchmark is available.

Taken together, these studies motivate the development of HalluTCG-Bench as a dedicated evaluation artifact for hallucination in LLM-generated test cases. By grounding evaluation in real requirements, standardizing annotations, and enabling cross-model comparison, HalluTCG-Bench would directly address Gap G1 and provide the empirical infrastructure needed to assess and improve hallucination behavior in NL-to-TCG pipelines.

### *B. R2 - Architect a Requirement-to-Test Traceability Pipeline (addresses G2)*

This recommendation addresses Gap G2, which identified the lack of end-to-end, auditable traceability between natural language requirements and generated test cases in fully automated NL-to-TCG systems. As established in Section VII, existing approaches achieve traceability only under restrictive conditions or as partial artifacts, while fully generative LLM-based systems produce no requirements to test traceability artifacts at all.

The partial traceability provided by CiRA [11], in which each generated test is anchored to an extracted conditional structure, represents the closest existing starting point within NLP-based systems operating on unconstrained natural language. This observation motivates the design of a provenance-tracking post-processing pipeline that can generate an auditable requirements-to-test traceability matrix (RTTM) alongside the test cases it generates, rather than treating traceability as an external or manual activity.

Architectural precedents demonstrating that such traceability is feasible already exist beyond end-to-end NL-to-TCG systems. Sawada et al. [30] demonstrate operational NLP + ML-based requirements to test traceability at NASA JPL, while Ge et al. [55] (IEEE TSE 2025) show that large language models can be used for cross-level requirements tracing. In parallel, UMTG [10] provides a validated reference architecture for formal constraint extraction that leverages OCL-based representations.

R2 therefore proposes the development of an auditable RTTM pipeline that integrates provenance tracking with test case generation, building on CiRA's conditional anchoring [11], formal constraint modeling as exemplified by UMTG [10], and LLM-assisted traceability techniques demonstrated by Sawada et al. [30] and Ge et al. [55]. The objective is not to replace existing generation pipelines, but to complement them with an explicit, inspectable traceability artifact that preserves automation while restoring auditability.

The adoption of such an RTTM pipeline would directly address Gap G2 by reconciling full automation with formal traceability, enabling NL-to-TCG systems to satisfy core requirements for verification, certification, and accountability in industrial and safety-critical contexts.

### *C. R3 - Build a Complexity-Stratified Benchmark (addresses G3)*

This recommendation addresses Gap G3, which identified systematic accuracy degradation as requirement complexity increases and the absence of standardized, complexity-aware evaluation frameworks for NL-to-TCG systems. As shown in Section VII, both traditional NLP-based approaches and LLM-based systems exhibit sharp performance drops when moving from simple to multi-condition requirements. Yet, existing evaluations remain fragmented, domain-limited, and difficult to compare across studies.

R3 proposes the construction of a multi-domain, complexity-stratified benchmark for natural-language-to-test-case generation. The benchmark would combine the PURE dataset [54], consisting of 79 software requirements specification documents and 34,268 sentences, with the real satellite Functional Requirement Documents (FRDs) used by Shakthi et al. [16], the user story corpora from SPECMATE [9], and the industrial requirements employed in CiRA [11], which are now available with ground-truth test artifacts. Together, these sources serve as the initial building blocks and provide coverage across multiple domains and levels of structural complexity. The benchmark can be continuously enhanced with additional data.

By explicitly stratifying requirements by complexity (e.g., number of conditions, signals, or interacting constraints), the benchmark would enable consistent evaluation of how accuracy degrades as requirements become more complex, thereby addressing the reproducibility and comparability limitations identified in current studies. In addition, Garousi et al. [27] identify further annotated requirements corpora that could be incorporated to extend domain coverage and support broader empirical validation.

The development of such a benchmark would directly address Gap G3 by providing a shared, publicly available evaluation framework that supports cross-domain comparison, complexity-aware analysis, and reproducible benchmarking of NL-to-TCG approaches. This, in turn, would enable clearer assessment of progress across techniques and architectures and support more reliable conclusions about their scalability and robustness (VIII C).

### D. R4 - Regulation-Compliant On-Premises Deployment Blueprints (addresses G4)

This recommendation addresses Gap G4, which identified unresolved ethical, security, and regulatory constraints that limit the adoption of AI-based test case generation in safety-critical and enterprise environments. As shown in Section VII, only two primary studies in the corpus - Yin et al.'s on-premises LLaMA-3 deployment [14], which is tabulated in Table IV, and Shakthi et al.'s ISRO satellite-FRD deployment [16], discussed in Section V, explicitly address data residency, export-control sensitivity, and deployable model-size constraints. Even in these cases, broad domain applicability and formal traceability remain unaddressed.

R4 proposes developing regulation-compliant, on-premises deployment blueprints for NL-to-TCG systems. These blueprints are intended to document architectural and operational patterns for deploying test case generation pipelines entirely within controlled enterprise environments, without reliance on external APIs or cloud-hosted inference endpoints.

Empirical evidence demonstrates both the feasibility and the limitations of such deployments. Yin et al. [14] report that a 4-bit quantized LLaMA3 70B system can achieve up to 90.3 percent coverage under on-premises constraints, representing the highest reported empirical ceiling in Table IV for an enterprise deployable configuration. Shakthi et al. [16] further show that open-source LLMs enable secure in-house processing of export-controlled satellite requirements. However, neither study provides guidance on traceability artifacts, compliance documentation, or generalization beyond the deployment context.

The proposed deployment blueprints would consolidate such operational evidence into a reusable reference framework that outlines how NL-to-TCG systems can be configured, audited, and validated under regulatory and confidentiality constraints. The adoption of these blueprints would directly address Gap G4 by enabling the secure, compliant, and inspectable deployment of AI-based test-case-generation systems in regulated domains, thereby supporting responsible industrial adoption (VIII-D).

### E. Emerging Trends

Beyond the four guidelines above, five emerging trends are shaping the next decade of intelligent software testing. Hybrid symbolic-neural architectures combine rule-based engines with the semantic reasoning capabilities of large language models (LLMs), allowing symbolic components to encode domain-specific rules or compliance constraints, while LLMs generate candidate test cases and explanatory rationales [34]. Multimodal requirement understanding extends test case generation beyond text by integrating information from natural language descriptions, UML/GUI diagrams, and code artifacts, enabling more comprehensive end-to-end test synthesis.

At the model-adaptation level, self-adapting LLMs incorporate continuous fine-tuning through reinforcement learning driven by test-execution feedback, often complemented by active-learning strategies to validate ambiguous or low-confidence cases [20]. From an assurance perspective, explainable AI for testing emphasizes traceable, transparent rationales that link requirements to generated test scripts, an essential property for auditability and regulatory compliance in safety-critical domains [22]. Finally, integration with DevOps pipelines enables automated, real-time regression testing and continuous integration/continuous delivery; however, modular AI components that interoperate cleanly with popular CI/CD tools and standardized APIs remain an active area of maturation [35].

### F. Closing Synthesis

Taken together, the analysis of twenty-six primary studies, the evidence-based identification of four structural research gaps (G1-G4), and the corresponding recommendations (R1-R4) provide a coherent roadmap for advancing AI-based test case generation from natural language requirements. The proposed benchmarks, traceability pipelines, and deployment blueprints address foundational limitations in hallucination control, traceability, complexity sensitivity, and regulatory compliance. At the same time, the emerging trends outlined in Subsection E - hybrid symbolic-neural architectures, multimodal requirement understanding, self-adapting LLMs, explainable AI, and DevOps integration indicate how these solutions may evolve into more robust, transparent, and industrially deployable testing ecosystems. Together, these findings answer RQ1-RQ4 by clarifying the state of the art and identifying concrete, evidence-driven directions for addressing its most consequential limitations.

The following section concludes the survey by summarizing its principal findings, revisiting the research questions, and outlining the implications for future research and practice.

## IX. Conclusions

Automated test case generation from natural language requirements has evolved from early rule-based parsing and deterministic grammars to contemporary large-language-model-based semantic reasoning. Foundational frameworks such as NAT2TESTSCR [3] and UMTG [10] established a critical bridge between linguistic analysis and formal testing by showing that natural language inputs can be translated into analyzable test artifacts. Among the surveyed systems, UMTG provides the most complete hybrid symbolic-neural realization, achieving 95 percent correct OCL constraint generation across two industrial automotive case studies.

Building on this foundation, domain-specific LLM systems such as LLM4Fin [13] demonstrate that explicit domain knowledge can substantially improve performance in regulated settings, achieving a peak business-scenario coverage of 98.18 percent and a five-dataset average of 91.89 percent, compared with ChatGPT's 49.65 percent average. In parallel, the on-premises deployment of quantized LLMs [14] demonstrates that high-coverage generation is feasible in security-sensitive environments, although rigorous benchmarking of such constrained deployments remains limited.

This survey identified twenty-six primary studies on test case generation from natural language software requirements published between 2000 and 2025, following the strict application of inclusion criteria I4 and I5 and venue-quality criterion E6.

Despite sustained progress, major challenges remain, including hallucinations in generated test cases, limited explainability and formal traceability, data scarcity for minority domains, and the absence of standardized evaluation benchmarks. Addressing these limitations will require balanced integration of hybrid symbolic-neural architectures, domain-knowledge grounding, and responsible AI testing practices. The four evidence-driven recommendations proposed in this survey - HalluTCG-Bench (R1), the requirements-to-test traceability matrix/pipeline - RTTM (R2), the complexity-stratified benchmark (R3), and the regulation-compliant deployment blueprint (R4) together provide a coherent roadmap for addressing structural gaps G1-G4 and advancing trustworthy, deployable automation for requirements-based software testing.

Overall, this paper provides an evidence-based survey that synthesizes twenty-six primary studies into a comparative framework, exposes the field's central trade-offs, and derives four concrete research recommendations for advancing trustworthy, scalable, and industrially deployable AI-based test case generation from natural language requirements.

## ACKNOWLEDGMENTS

The author expresses sincere gratitude to the researchers whose work underpins this survey. Their foundational contributions to requirements-based testing, natural language processing, and AI-driven software testing have been essential in advancing the state of the art examined in this paper. The author also thanks colleagues and reviewers for their constructive feedback, which helped improve the clarity, rigor, and organization of this study.

The author is particularly grateful to Dr. Yetunde Orimoloye for her valuable input during the review process. Her insightful feedback helped refine the research approach, and her expertise was instrumental in supporting data analysis and ensuring the accuracy and rigor of the results presented.